\documentclass[preprint,10pt]{elsarticle}
\usepackage{amssymb}
\usepackage{array} 
\usepackage{caption} 
\usepackage{hyperref}
\hypersetup{hidelinks} 
\usepackage{booktabs}
\usepackage{graphicx}
\usepackage{subfigure}
\usepackage[section]{placeins}
\usepackage[ruled]{algorithm2e}
\usepackage[table]{xcolor}
\usepackage[margin=0.8in]{geometry} 

\newcolumntype{M}[1]{>{\arraybackslash}m{#1}}
\newcolumntype{O}[1]{>{\centering\arraybackslash}m{#1}}

\journal{arXiv}
\begin{document}

\begin{frontmatter}
\title{Modeling of Memory Mechanisms in Cerebral Cortex and Simulation of Storage Performance}
\author[]{Hui Wei}\corref{cor1}
\ead{weihui@fudan.edu.cn}
\author[]{Chenyue Feng}
\author[]{Jianning Zhang}
\cortext[cor1]{Corresponding author}
\affiliation[]{organization={Laboratory of Algorithms for Cognitive Models, School of Computer Science, Fudan University},
            addressline={No.2005 Songhu Road}, 
            postcode={200438}, 
            city={Shanghai},
            country={China}}
\begin{abstract}
At the intersection of computation and cognitive science, graph theory is utilized as a formalized description of complex relationships and structures. Traditional graph models are often static, lacking dynamic and autonomous behavioral patterns. They rely on algorithms with a global view, significantly differing from biological neural networks, in which, to simulate information storage and retrieval processes, the limitations of centralized algorithms must be overcome. This study introduces a directed graph model that equips each node with adaptive learning and decision-making capabilities, thereby facilitating decentralized dynamic information storage and modeling and simulation of the brain’s memory process. We abstract different storage instances as directed graph paths, transforming the storage of information into the assignment, discrimination, and extraction of different paths. To address writing and reading challenges, each node has a personalized adaptive learning ability. A storage algorithm without a “God’s eye” view is developed, where each node uses its limited neighborhood information to facilitate the extension, formation, solidification, and awakening of directed graph paths, achieving competitive, reciprocal, and sustainable utilization of limited resources. Storage behavior occurs in each node, with adaptive learning behaviors of nodes concretized in a microcircuit centered around a variable resistor, simulating the electrophysiological behavior of neurons. Based on Ohm’s and Kirchhoff’s laws, we simulated the dynamics of this directed graph network on a computer, where the network could store and retrieve uploaded instances, confirming the model’s effectiveness and exploring its storage capacity. Under the constraints of neurobiology on the anatomy and electrophysiology of biological neural networks, this model offers a plausible explanation for the mechanism of memory realization, providing a comprehensive, system-level experimental validation of the memory trace theory.
\end{abstract}

\begin{keyword}
 Memory Modeling\sep Directed Graph\sep Storage Simulation\sep Decentralized Algorithm
\end{keyword}
\end{frontmatter}
\section{Introduction}
The brain, as a complex information processing system, remains incompletely understood, particularly its memory mechanisms. Exploring how the brain encodes, stores, and retrieves information has always been a significant challenge in memory research. Modern neuroimaging technology is rapidly advancing, and through methods such as fluorescent tagging and two-photon imaging, humans are gradually unveiling the neural system’s connectivity and other details. However, such analyses can only offer limited insights, and we are still unfamiliar with the neural mechanisms and information flow directions underlying memory activities. From a neurobiological perspective, memory refers to neural system activities or physical changes in neuronal connections triggered by external stimuli or brain states \cite{chaudhuri_computational_2016}. Neurobiological research on memory mechanisms focuses on synaptic plasticity \cite{martin_synaptic_2000,abraham_is_2019,jeong_synaptic_2021,humeau_next_2019}, neurotransmitters \cite{de_rossi_neuronal_2020,feld_neurochemical_2020}, and electrochemical signals \cite{wang_single-vesicle_2022,otero_exploring_2022} at the cellular and molecular levels, while cognitive psychology likens memory to an information processing system responsible for encoding, storing, and retrieving information. These theories provide a reasonable starting point to map the brain’s coarse-scale organization using functional imaging technologies including EEG and fMRI \cite{kriegeskorte_cognitive_2018}.

\begin{figure}[htbp]
    \centering
    \includegraphics[width=10cm]{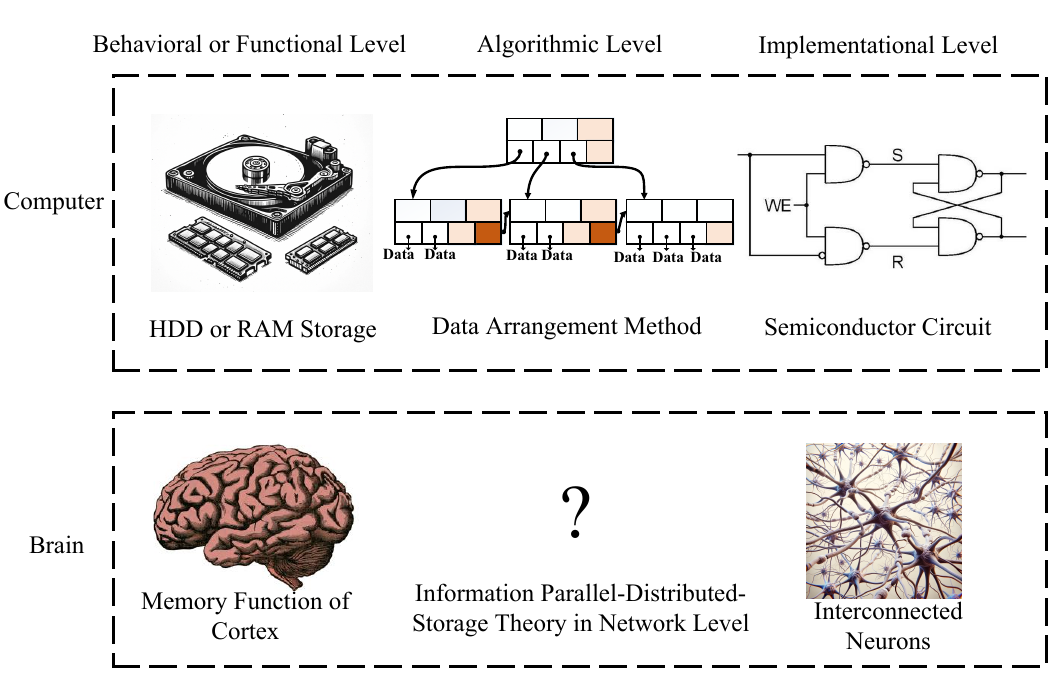}
    \caption{Marr’s Hierarchical Diagram of Storage Memory Systems of Computers and the Brain}
    \label{fig:marr}
\end{figure}

Marr’s three-level theory divides the problem to be studied into three levels: behavioral, algorithmic, and implementation. We adopt this theory to stratify the study of memory mechanisms, comparing the storage functions of computers and the human brain in a layered manner, as illustrated in Figure \ref{fig:marr}. Current understanding of the brain’s algorithmic layer is relatively limited, especially intermediate models that are consistent with neurobiological experimental evidence. Therefore, it is necessary to combine computational theory with biological experiments, modeling and simulating the brain’s memory process based on neurobiological evidence, bridging the gap between the microscopic level of cells and the macroscopic level of cognitive behavior.

We model and simulate the biological neural network to explore the brain’s memory mechanism, focusing on how encoding, storage, consolidation, and retrieval of information occur in the brain’s memory process. Modern digital computers possess powerful storage capabilities, and comparing them with the brain can further our understanding of what is still missing in memory research. Table \ref{tab:comparison} compares computer storage and brain memory management, from which the implementation process of computer storage, from the bottom to the top layer, is clear. However, the implementation process of brain memory involves numerous unknowns, which motivates us to simulate and emulate the brain’s memory process based on neurobiological mechanisms.

\begin{table}[htbp]
\setlength{\arrayrulewidth}{0.2pt}
\centering
\footnotesize
\caption{Comparison of Computer Storage Management and Brain Memory Management}
\label{tab:comparison}
\begin{tabular}{O{2cm}|O{6cm}|O{6cm}} 
\toprule 
\textbf{Category} & \textbf{Computer Storage Management} & \textbf{Brain Memory Management} \\
\midrule
\end{tabular}

\begin{tabular}{O{2cm}|M{6cm}|M{6cm}} 
Minimum Functional Unit & Data are stored in binary form, with hard drives using polarity of magnetic particles, charges in capacitors, etc., to represent 1s or 0s. & In neural cells, membrane possesses resting and action potentials. In resting state, signal is inhibited; if activated, it generates an action potential, conducting signal. \\
\midrule
Storage Device & An HDD records data by changing the distribution of the two polarities of magnetic particles using an electromagnet on read/write head; an SDD stores 0s and 1s by altering the number of electrons in the floating gate layer through the application of an electric field. & Memory refers to neural system activities or physical changes in neuronal connections triggered by external stimuli or brain states \cite{chaudhuri_computational_2016}. Many studies have begun to define engrams as basic units of memory, but there are unresolved issues such as how structure of engrams affects quality of memory, how multiple engrams interact with each other, and how engrams change over time \cite{josselyn_memory_2020}.\\
\midrule
Data Allocation & Operating systems allocate disk space for files with blocks as the basic unit, using methods such as contiguous allocation, linked allocation, and indexed allocation. & We have a very limited understanding of how a single memory item is structurally stored at the level of biological neurons, and how multiple memory items are allocated among neural cells (groups) in cerebral cortex. \\
\midrule
Data Access & Control circuits translate logical addresses into physical addresses. Read/write head of an HDD moves to the cylinder, track, and sector where the data are located for data access; for an SDD, controller directly performs read and write operations on storage units at physical address. & When brain stores or recalls a memory, we know little about how it locates engram of memory content among vast number of neural cells in cerebral cortex, replacement of old memories with new ones, and neuronal network mechanisms during recall. \\
\bottomrule
\end{tabular}
\end{table}

We focus on the algorithmic layer of the brain’s memory system, aiming to model and simulate the memory processes of the brain. We seek to explain the mechanism of memory realization from the algorithmic layer, exploring how the biological neural network encodes, stores, consolidates, and retrieves information. This contributes to the brain's memory mechanisms and investigates their alignment with neurobiological experimental evidence.

The brain can be conveniently represented as a network of neurons and their interconnections, making graph theory a mathematical tool for studying its structure and functional systems. We utilize directed graphs from graph theory to model and simulate the algorithmic layer of the brain’s memory system, aiming to align anatomical and physiological data. As a branch of mathematics, graph theory provides tools for processing and analyzing network structures. Tracing back to Euler’s solution to the Seven Bridges of Konigsberg problem in the 18th century \cite{gao_graph_2009}, it has been applied such as to design solutions for the Traveling Salesman Problem(TSP) \cite{cook_traveling_2007}, construct knowledge graphs \cite{ji_survey_2021}, and create databases using graph structures \cite{jouili_empirical_2013}, where graphs usually serve as structured representations of data or knowledge, inherently lacking dynamic behavior, with their functionality reliant on externally applied algorithms.

\begin{figure}[htbp]
    \centering
    \subfigure[Centralized Mode: Single Global Module Operates on Graph from Detached, Overarching Perspective]{\includegraphics[width=7cm]{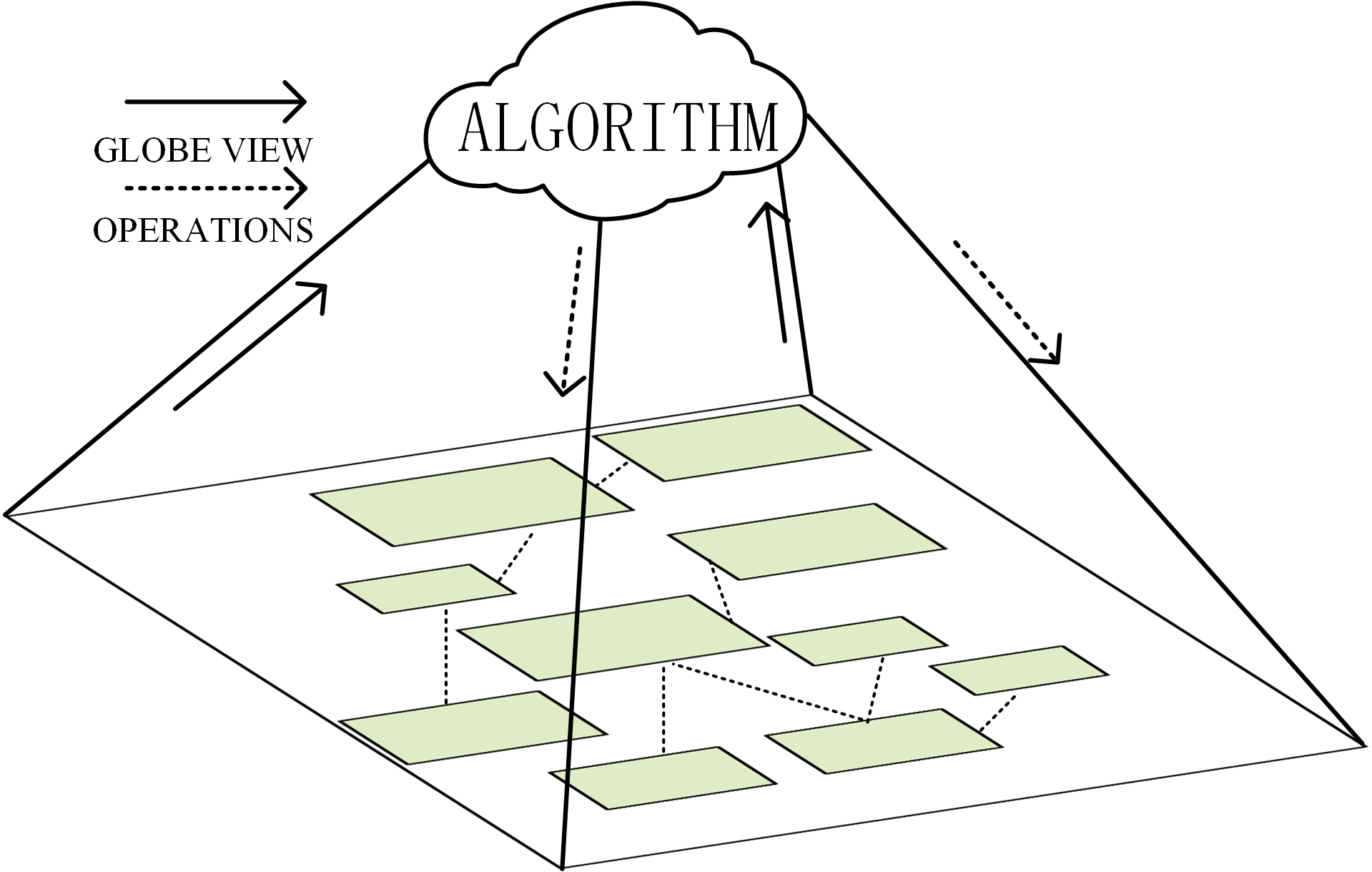}}
    \subfigure[Decentralized Mode: Each Node Possesses Attributes and Autonomous Capabilities, Such as Establishing Connections with Other Nodes]{\includegraphics[width=7cm]{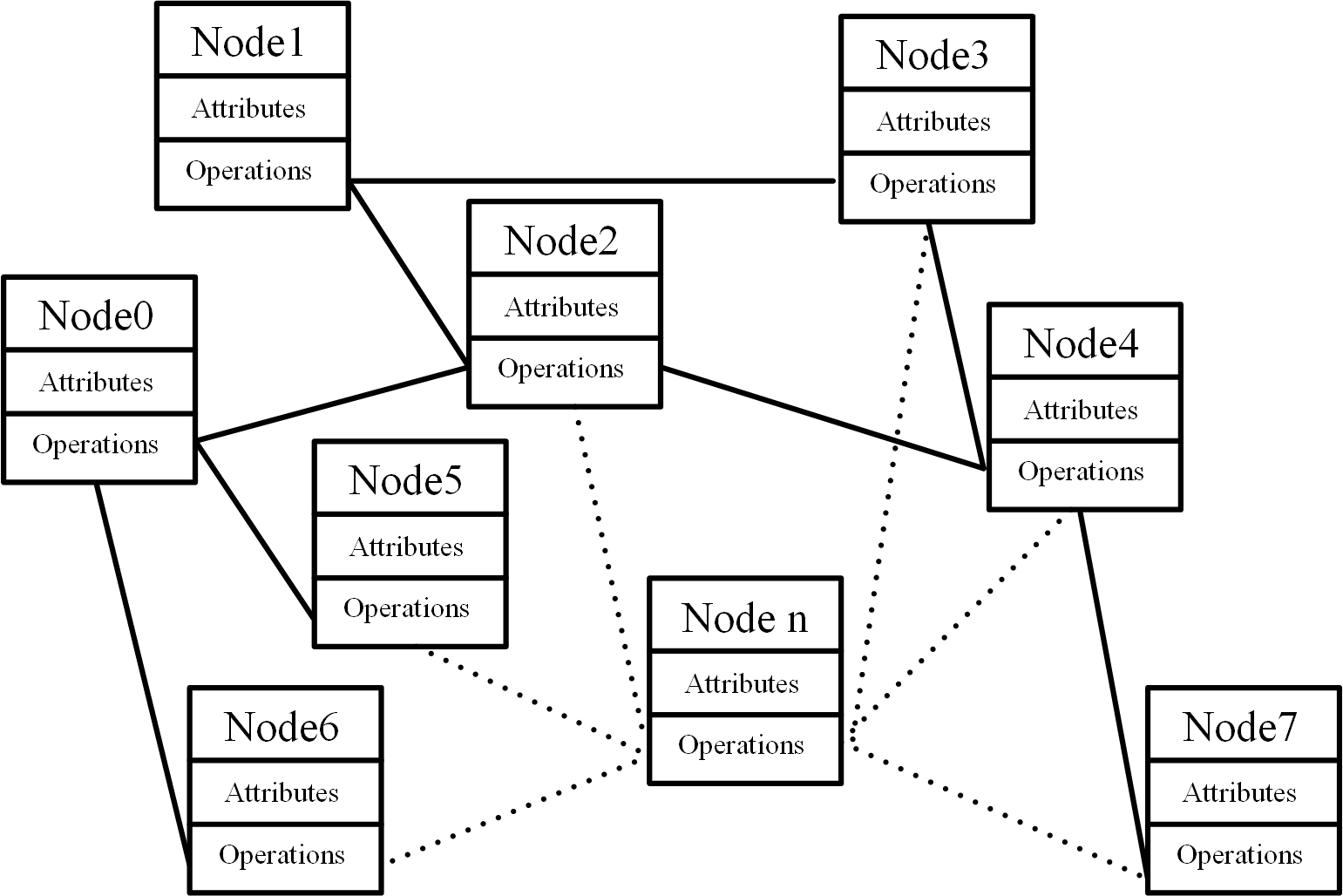}}
    \caption{Comparison of Centralized and Decentralized Modes}
    \label{fig:compare}
\end{figure}

In these traditional applications of graph theory, algorithms typically operate on the graph structure from a global view, meaning that the executor of the algorithm (like a CPU) has access to global information and makes decisions. While biological neural networks lack such a global view or central controller and are characterized by decentralization, consisting of many simple units that are only connected to their neighbors. To better align with biological neural networks and more accurately model and simulate biological memory processes, we propose an algorithm that does not rely on a “God’s eye” view, focusing on implementing a storage function in a decentralized graph model. Just as neurons have only local connections, nodes in a directed graph can only see their connected neighbors. Each node makes decisions based on its local field of information. Nodes are no longer passive data storage units, but are active, autonomous units, which can adaptively learn how to respond in different contexts, simulating the behavior of biological neurons. Passive nodes with a single global algorithm and active nodes with numerous independent small algorithms represent two completely different paradigms. Contrasting these two modes in Figure \ref{fig:compare}, we see the difference between traditional centralized processing and the proposed decentralized processing. In the latter, each node stores information and can also process and transmit information, forming complex dynamic patterns across the entire network.

We focus on the information processing mechanisms behind brain memory activities, using graph theory to comprehensively model and simulate memory, abstracting storage instances into directed graph paths. Through decentralized directed graphs and the pervasive learning algorithms inherent in each node, it dynamically learns and optimizes these paths, achieving a memory mechanism based on directed graphs, enabling efficient information storage and retrieval, and offering a novel perspective on the brain’s memory mechanisms.

\section{Common Methods for Modeling Parallel Distributed Processes}
The realization of memory functions in the biological cerebral cortex is a parallel distributed process, which can be modeled by several common methods.
\subsection{Classical Graph Theory}
Graph theory is widely applied in computer science, with current research focusing on the properties of static graphs. However, it relies on a “God’s eye” view and lacks research into the self-organizing dynamic behavior of graphs.

The TSP is a classic NP-hard problem, which involves finding the shortest path that visits each node exactly once and returns to the starting point in a complete graph \cite{cook_traveling_2007}. It has a wide impact in areas including logistics, flight route planning, network design, and bioinformatics. Due to its high computational complexity, various methods have been proposed to solve or approximate it, such as branch and bound, nearest neighbor, simulated annealing, and genetic algorithms. These methods essentially act on a completely known graph, possessing a “God’s eye” view.

One of the earliest large-scale applications of classical graph theory in computer science is graph databases. Representing modern applications such as social networks in graph form is quite natural  \cite{jouili_empirical_2013}. Graph databases represent data or patterns as graphs, with data structures modeled as directed, possibly labeled graphs. Data operations are represented by graphs, and operation-oriented and type constructors can be defined on graph structures along with appropriate integrity constraints \cite{angles_survey_2008}. As defined, the implementation of graph databases relies on external algorithms for different operations, such as graph construction, exhibiting a “God’s eye” view.

Another important application of classical graph theory is the knowledge graph, i.e., a structured representation of facts, entities, relationships, and semantic descriptions. Knowledge is interconnected, with knowledge nodes forming a graph network to integrate various knowledge domains \cite{ji_survey_2021}. Knowledge graphs possess a complete topology and leverage their structure to integrate knowledge, providing a “God’s eye” view. However, the graph remains static after information updates.

\subsection{Graph Theory and Deep Learning}
Graphs are now widely used in deep learning. Tasks involving models on graphs can categorized into two types: Node-centered tasks are associated with individual nodes in a graph and include node classification, link prediction, and node recommendation. Graph-centered tasks are related to the entire graph and include graph classification, estimation of various graph properties, and graph generation \cite{zhang_deep_2020}. A well-known application, the Graph Convolutional Network (GCN), inputs graphs to convolutional networks for training \cite{jia_multimodal_2023}. Operations in deep learning involve manipulating graph data, relying on a “God’s eye” view, while the graph remains static.

Graph theory is extensively used in deep learning applications. In computer vision, node encoding methods include both non-neural networks, such as traditional graph embedding and probabilistic graph models, and neural networks, including Graph Recurrent Neural Networks, GCNs, and variants of Graph Neural Networks \cite{jiao_graph_2022}. Again, these methods process graph data, and the graph itself is static.

\subsection{Multi-agent Systems}
Multi-agent systems have gained widespread attention in recent years. These decentralized systems accomplish a task through the collaborative effort of many small entities, each with independent behavior. The small entities make distributed independent decisions with individual tasks assigned to autonomous entities called agents, each determining the correct action for solving the task based on their inputs. Individual agents only have partial information and communicate with their neighbors. These are applied in areas such as computer networks, drones, and robot swarms \cite{dorri_multi-agent_2018}.

Multi-agent systems are inspired by the natural world, such as in ant colonies and bee societies \cite{leitao_bio-inspired_2012}. Each ant in a colony achieves complex tasks such as finding food, building nests, and collective defense through simple pheromone communication and collaborative behavior. Each bee communicates effective foraging information, temperature regulation, and new nest site selection through precise dances and collaboration. Through local interactions and simple rules, they achieve complex global behavior, providing inspiration and reference for computer science and promoting the development and innovation of artificial multi-agent systems.

Multi-agent systems are extensively used in robotic systems and to some extent can solve issues such as local positioning, obstacle detection, path planning, and navigation in multi-robot systems \cite{rasheed_review_2022}. For example, a collision avoidance problem was solved for multiple robots in a decentralized, distributed framework, where input data were only collected from onboard sensors \cite{long_towards_2018}. Each robot is an agent that shares its decision with all other robots, thereby achieving an optimal strategy among them.

Although multi-agent systems utilize distributed and decentralized decision-making, the connections between agents are relatively weak and do not form a fixed pattern for storage or other applications.

\section{Modeling Based on Directed Graphs}
Many researchers have modeled memory. The Hopfield network \cite{hopfield_neural_1982} is a type of recurrent neural network that can be abstracted as a directed complete graph, with many variants \cite{krotov_dense_2016,kobayashi_chaotic_2017,oku_pseudo-orthogonalization_2013,mungai_chunking_2017,mungai_semantic_2017,shriwas_multi-modal_2019}. Bidirectional associative memory networks \cite{kosko_bidirectional_1988} implement bidirectional associative functionality and have undergone many modifications \cite{zhao_synchronization_2021,kosko_bidirectional_2021,singh_multilayer_2017,cholet_bidirectional_2019}. Structural learning has also been widely used \cite{gershman_computational_2017,wei_hierarchical_2007,kimoto_mixed_2004,dekhtyarenko_systematic_2005,he_constructing_2019}, with some models considering the impact of memory retrieval on memory storage \cite{gershman_computational_2017}. There are also sparse associative memory models \cite{hoffmann_sparse_2019} and models based on dictionary learning \cite{mazumdar_associative_2017,tosic_dictionary_2011}. Models based on graph theory, such as clique-based models \cite{gripon_sparse_2011} and competition graph models \cite{jiang_storing_2015,mofrad_neural_2021}, use graph theory and network flow theory to simulate human memory. These models have made progress in simulating human memory, but nevertheless have limitations. For instance, dense associative memory\cite{krotov_dense_2016} lacks sufficient simulation of neuronal structure and function; the Hopfield \cite{hopfield_neural_1982} and BAM networks \cite{kosko_bidirectional_1988} have a gap in fully connected networks and biological realism; and clique-based models \cite{gripon_sparse_2011} process memory in a way not entirely consistent with actual mechanisms in the human brain. Many models face challenges in hardware implementation, computational resource requirements, and data processing efficiency.

Electronic components such as memristors \cite{chua_memristor-missing_1971,tour_fourth_2008,itoh_memristor_2008} have also been proposed to model memory. Their resistance changes when electric current passes through, but remains constant if current flow stops. This memory of resistance makes them like biological neural synapses, and they are often used to construct memory computing models \cite{sun_memristor-based_2019,sun_quasi-ideal_2014,an_realizing_2019,li_novel_2019,chen_adaptive_2018}. However, memristor-based memory models have shortcomings, such as only simulating a few neuronal synapses, and thus differing from biological reality \cite{sun_memristor-based_2019}, or focusing on circuit analysis and neglecting biological emulation \cite{chen_adaptive_2018}.

\begin{figure}[htb]
    \centering
    \includegraphics[width=12cm]{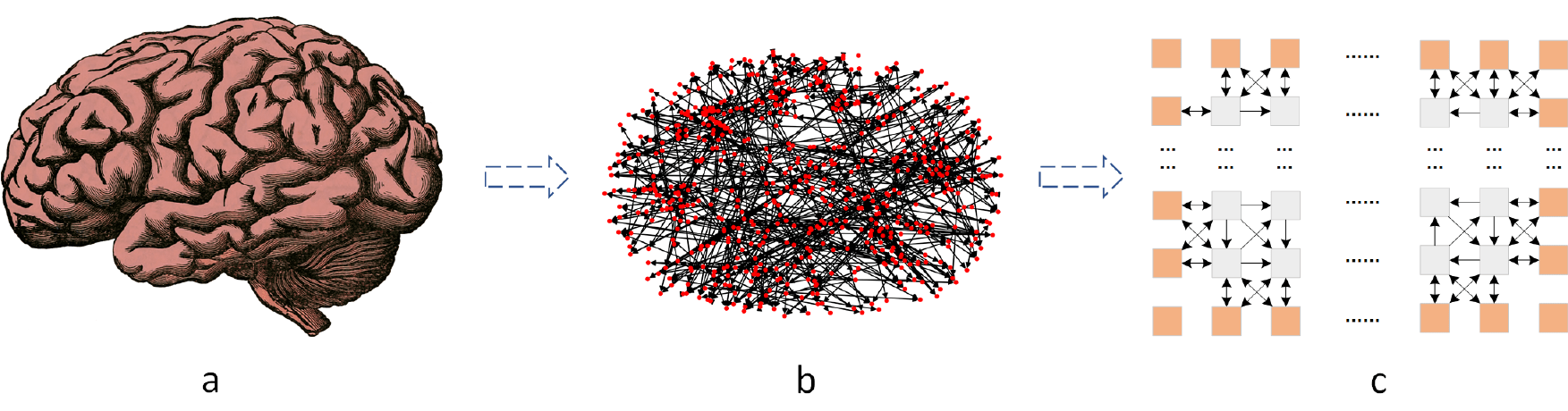}
    \caption{Abstraction of Human Brain Memory System into a Locally Connected Directed Graph}
    \label{fig:b2g}
\end{figure}

These models provide valuable perspectives and tools to understand and simulate human memory. Based on them, our proposed model can more accurately simulate and understand the complex memory processes in the human brain, whose memory system consists of many structurally similar neurons connected by synapses, which can be abstracted as nodes and edges in a graph network, as shown in Figure \ref{fig:b2g}. Abstractly modeling this as a locally connected directed graph aligns with biological reality. The process of signal transmission between neurons can be abstracted as a dynamic network flow.

\begin{figure}[htbp]
    \centering
    \subfigure[]{
    \includegraphics[height=4cm]{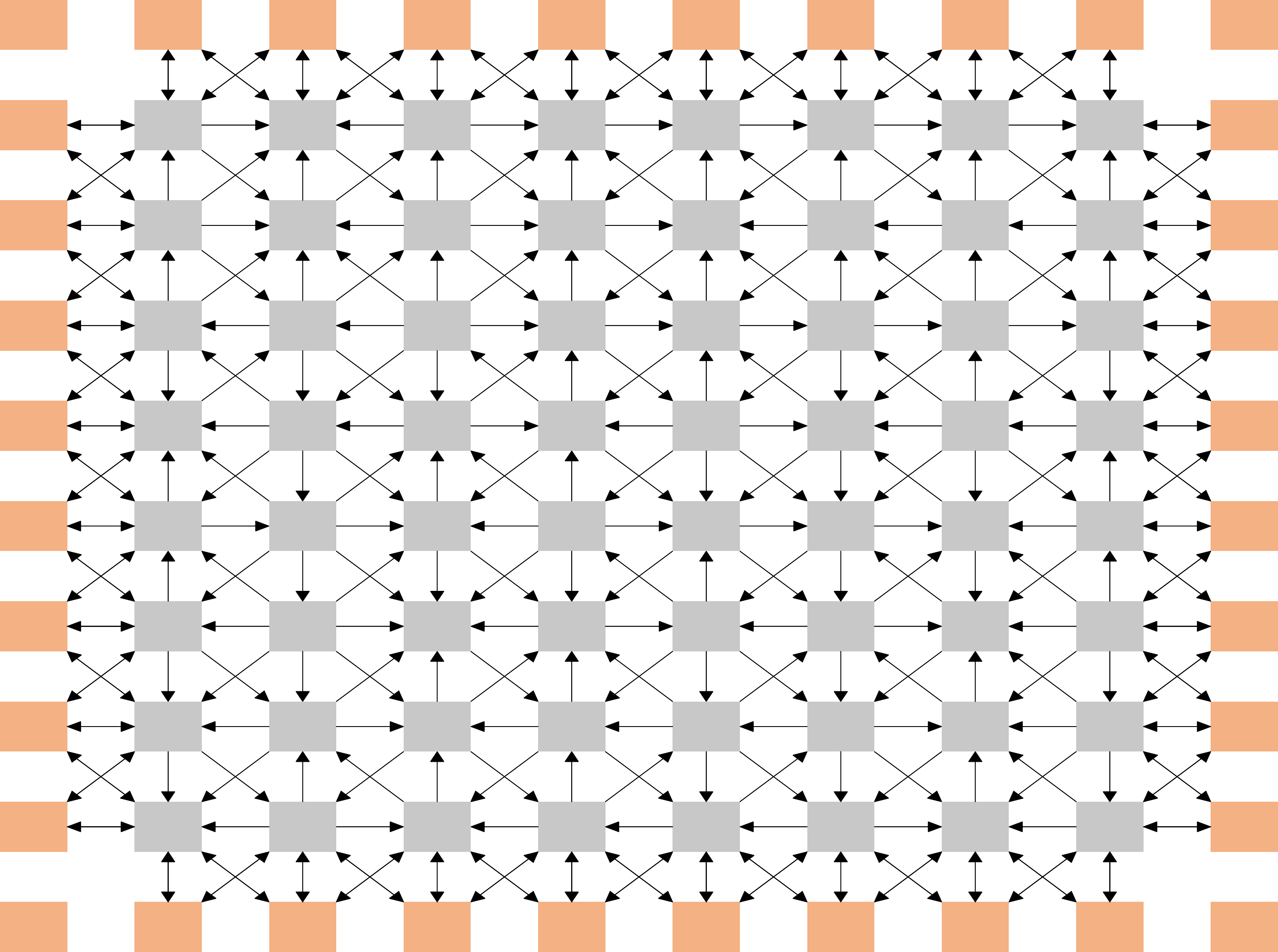}
    \label{fig:base10}
    }
    \hspace{5mm}
    \subfigure[]{
    \includegraphics[height=4cm]{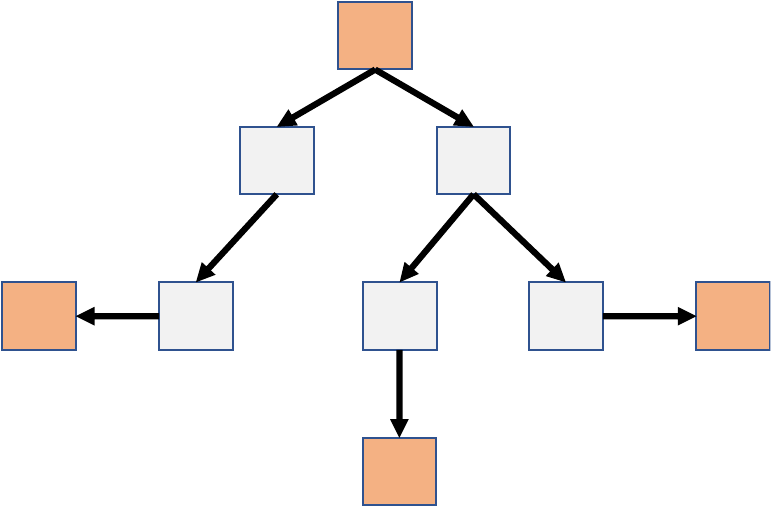}
    \label{fig:stobyrt}
    }
    \caption{(a):Abstract Modeling of Neuronal Network as Sparsely Connected Directed Graph (b):Directed Graph Path Modeling Representing Storage Instances}
\end{figure}

We utilize a two-dimensional matrix of directed graph nodes to abstractly model the biological neural network. Figure \ref{fig:base10} depicts a generated 10$\times$10 scale node matrix of a directed graph network. Nodes, which possess their own behavior, are connected to neighbor nodes through directed edges. Edge nodes and internal nodes are shown in orange and gray, respectively. Arrows between nodes represent connections and directions of signal transmission. Notably, although there are bidirectional connections between edge and internal nodes, only one direction is activated at a time. When an edge node’s connection is directed toward an internal node, it is used as an input, representing the abstract definition of a storage instance, such as of an object’s name. When a connection is directed toward an edge node, it acts as an output node, representing specific features of a storage instance such as color or smell. A non-activated connection indicates that the storage instance lacks a certain specific feature. The proposed model’s underlying directed graph network should be capable of performing storage functions across different scales and topological structures.

According to the memory trace theory, the content of memory is physically the propagation of neural excitation in a biological neural network caused by input stimuli, along with the reinforcement and awakening of propagation traces. The repeated occurrence of input stimuli can strengthen the connectivity between nodes. Inputs correspond to different propagation path traces, meaning that connected paths are specific physical carriers. This process does not require an overarching “God’s eye” view and is reasonable within the context of a biological neural network, which is massive and dynamic. Therefore, in our modeling, different paths are formed by arcs in the directed graph as the physical implementation of storage. For example, in Figure \ref{fig:stobyrt}, the nodes at the top, bottom, left, and right, connected by some intermediate nodes and directed edges, form a subgraph that can be considered a storage instance in the directed graph.

\section{Formation and Consolidation of Directed Graph Paths}
Consider the implementation process of memory as the formation and consolidation of paths between nodes in a directed graph network is a natural approximation. Humans perceive external information through multiple senses to understand the characteristics of an object. Neurons, upon receiving stimuli, transmit signals to different downstream neurons in the network, forming various transmission paths. Repeated stimulation solidifies these paths, creating a memory. When similar signals are received again, they are transmitted along these established paths, in a process known as path consolidation.

Based on this process, our model, as shown in Figure \ref{fig:base10}, classifies nodes into edge nodes and internal nodes. Edge nodes represent the characteristic information of storage instances, serving as inputs and outputs of information. An activated edge node indicates the presence of a feature. Activated edge nodes can be in an input or output state (i.e., the respective starting point and endpoint of signal transmission), while inactivated nodes cannot input or output signals. Input edge nodes can connect and transmit signals to output edge nodes through internal nodes. The signal enters from an input node and continuously propagates outward, forming different paths, and the transmission ends when it reaches an output node or there is no further path. As the signal can only be output through output nodes, a stable signal flow is formed only on the directed graph paths that reach the output nodes; these interconnected paths form the path traces of an instance on the directed graph. Longer paths take more time to reach the output node, and signals are transmitted through the shortest path, which is the physical realization of storage, i.e., the connected path of the storage instance or simply the storage instance path.

\begin{figure}[htbp]
    \centering
    \subfigure[]{
    \includegraphics[width=6cm]{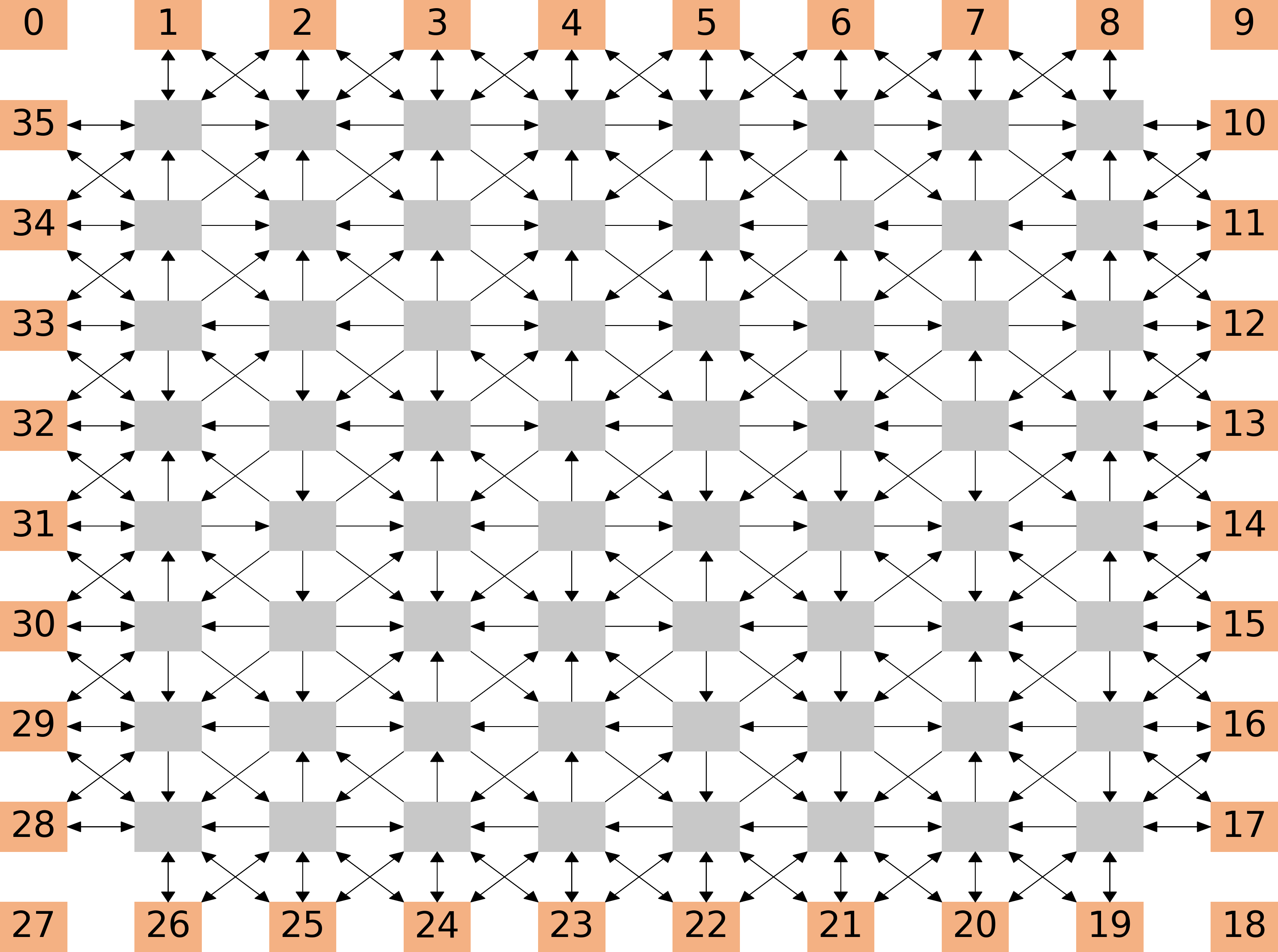}
    \label{fig:numedge}
    }
    \hspace{10mm}
    \subfigure[]{
    \includegraphics[width=6cm]{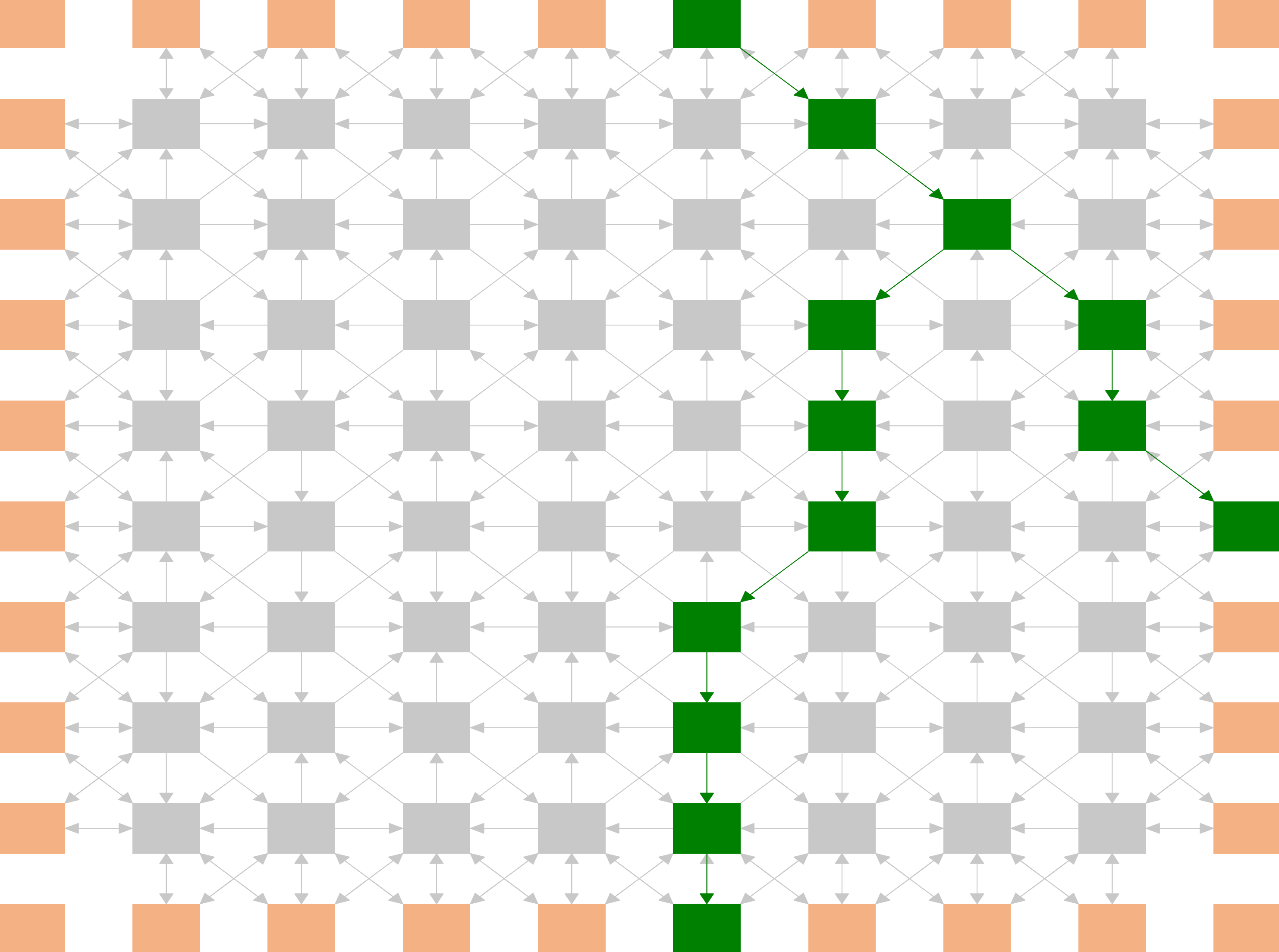}
    \label{fig:node2rt}
    }
    \caption{(a):Numbering of Edge Nodes in Directed Graph (b):Forming Paths by Changing States of Edge Nodes}
\end{figure}

Figure \ref{fig:numedge} shows the numbering of each edge node. A vector can be constructed by connecting the states of edge nodes according to their numbers. This can be seen as a storage instance vector, representing different external stimulus features by altering the vector’s bits, forming different interconnected storage instance paths that can be remembered by the directed graph. Table \ref{tab:v2state} shows a vector of edge node states, where the first and second rows show node numbers and node states, respectively, where 1 indicates a node activated in the input state, –1 indicates the output state, and 0 represents inactivation. This vector is a storage instance, and the green path in Figure \ref{fig:node2rt} represents the shortest transmission path of the signal at that time.

\begin{table}[htbp]
\setlength{\arrayrulewidth}{0.2pt}
\centering
\caption{Vector Representation of Edge Node State}
\label{tab:v2state}
\begin{tabular}{O{2cm}|c|c|c|c|c|c|c|c|c|c|c|c|c|c|c|c|c|c}
\toprule
\textbf{Edge Node} & 0  & 1  & 2  & 3  & 4  & 5  & 6  & 7  & 8  & 9  & 10 & 11 & 12 & 13 & 14 & 15 & 16 & 17 \\
\hline
\textbf{Status}    & 0  & 0  & 0  & 0  & 0  & 1  & 0  & 0  & 0  & 0  & 0  & 0  & 0  & 0  & -1 & 0  & 0  & 0  \\
\midrule
\textbf{Edge Node} & 18 & 19 & 20 & 21 & 22 & 23 & 24 & 25 & 26 & 27 & 28 & 29 & 30 & 31 & 32 & 33 & 34 & 35 \\
\hline
\textbf{Status}    & 0  & 0  & 0  & 0  & -1 & 0  & 0  & 0  & 0  & 0  & 0  & 0  & 0  & 0  & 0  & 0  & 0  & 0 \\
\bottomrule
\end{tabular}
\end{table}

Inspired by the complex regulation mechanisms of biological neuronal signal transmission, each node in the directed graph possesses independent and autonomous path learning capabilities, enabling nodes to record transmission paths. The directed graph can be divided into two working modes: storage and retrieval.

In the storage mode, the node’s path learning algorithm is activated. If there is a continuous flow of signals through a node, it will map the input path to the corresponding output path. If we upload the complete storage instance vector, activating the edge nodes, stable signal flow will be formed only on the shortest path. Once the path is formed, a stable and continuous signal flow will exist along it, and the path learning algorithm of each node will store the path that passes through, distributing the traces of the storage instance path across each node.

In retrieval mode, the node’s path learning algorithm is turned off. The vector uploaded during retrieval, i.e., the probe vector, may lose some bits compared with the original vector. When this is uploaded, edge nodes of the missing bits will be set to output status, allowing the activation of more output paths in the graph. This implies that in the absence of certain features, their possible existence is assumed, which is later determined by whether there is signal output, and the signal transmission path is the retrieved path. Notably, the input state bits cannot be lost, as otherwise signal transmission cannot be formed due to lack of input. In extreme cases, all bits except those in the input state are lost, in which case all other edge nodes are activated as output nodes, meaning that for each node, all possible output paths can be activated. At this time, nodes output signals according to the path traces stored during the storage process, activating and outputting signals through path traces previously formed in storage mode, thus achieving retrieval of the stored instance paths.

As can be seen from Figure \ref{fig:node2rt}, the paths in the directed graph network are a limited resource. Therefore, uploading multiple storage instances into one directed graph can result in competition for paths and node resources. When a path is occupied by one storage instance, other instances that also need this path can create resource competition, which can be incompatible and destructive. An ideal solution is to find a compatible path-sharing scheme for multiple storage instances. Then, during retrieval, additional features may be needed, not just the shared path for correct retrieval. This is like the human brain’s memory process, where the Hippocampus-Cortical two-tier system temporarily stores information. The hippocampus may undertake batch processing of different memory instances, forming an intensive, path-compatible, multi-instance composite representation to conserve resources, which is later separated and transferred to the cortex for long-term storage.

\section{Current Path Formation and Consolidation Algorithm Based on Electric Field}
In the neuronal network of the biological cerebral cortex, action potential pulses are transmitted along directed paths. Therefore, in our design, the directed graph network is endowed with the attributes of a circuit, using electric field theory to explain the establishment and consolidation of paths. This process is verified through computer simulation.
\subsection{Internal Structure Design of Nodes}
\begin{figure}[htbp]
    \centering
    \includegraphics[width=12cm]{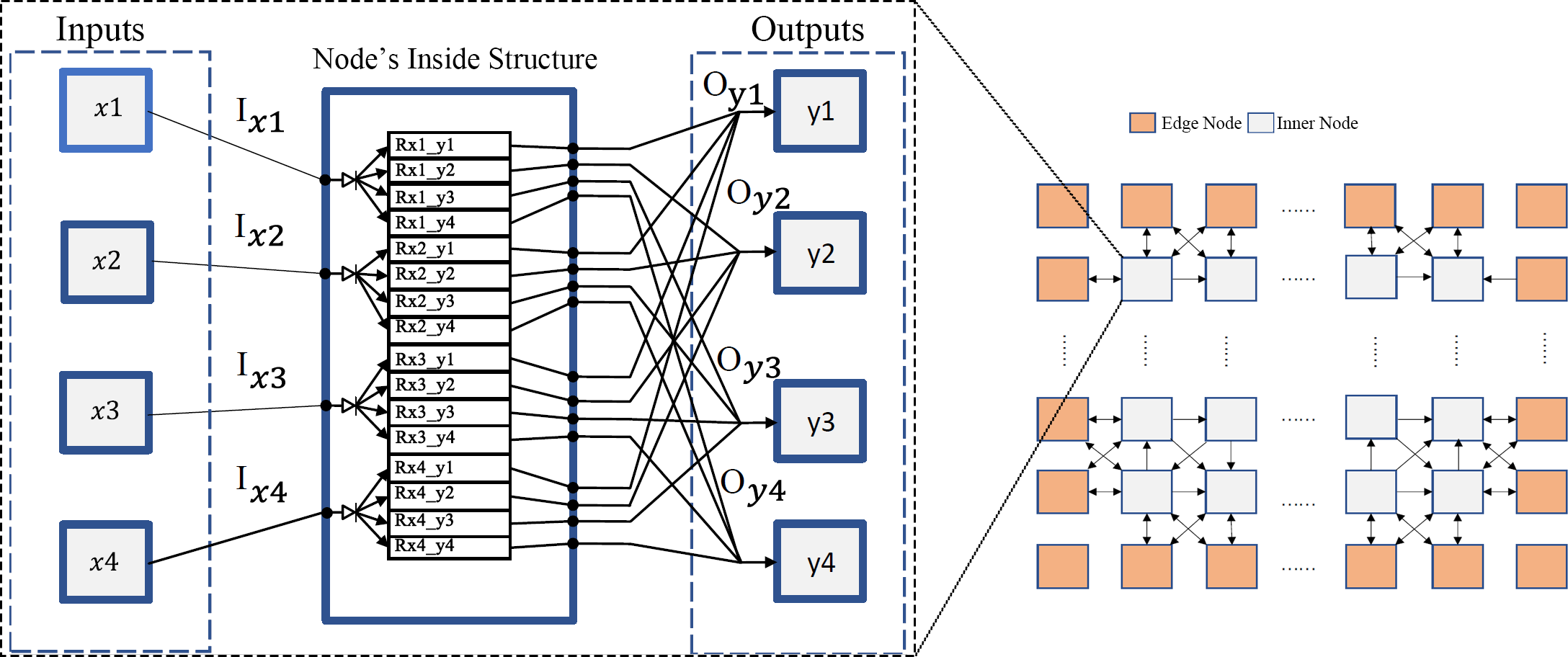}
    \caption{Modeling of Simulation Circuit Structure Inside Node of Directed Graph}
    \label{fig:innernode}
\end{figure}

A single biological neuron cell can differentiate various signal inputs and output different signals to various neighboring neurons. Based on this, we model a single node of the directed graph network, as shown in Figure \ref{fig:innernode}. There exists a path between each input and output, with diodes and variable resistors on each path. The total resistance value of the variable resistors is fixed, and they have collaborative change rules. Diodes ensure unidirectional current flow. These components enable path consolidation and resource competition learning between nodes.

The input and output of the node in Figure \ref{fig:innernode} is each connected to four nodes, where input nodes $x1, x2, x3, x4$ have respective incoming current values of $I_{x1}, I_{x2}, I_{x3}, I_{x4}$, and output nodes $y1, y2, y3, y4$ have respective outgoing current values of $O_{y1}, O_{y2}, O_{y3}, O_{y4}$. The variable resistors are $R_{x1\_y1}, R_{x1\_y2}, ..., R_{x4\_y4}$, representing resistances on different paths inside nodes from input to output. Variable resistors have the function of distributing the output current.

Edge nodes input by connecting to an external power source and output by grounding, or they can disconnect. Internal nodes receive input and output currents from neighboring nodes. Unlike internal nodes, the resistors inside edge nodes do not have collaborative change rules, and they only function to limit current and prevent short-circuiting.

\subsection{Voltage Application Method for Edge Nodes}
We model the signal transmission of the neural network as the flow of electric current in a circuit, in which current always flows from a higher to a lower voltage level. The paths in the directed graph are abstracted as current transmission paths between nodes. Voltage levels are set depending on the states of edge nodes. An edge node in the input state is set to high voltage and in the output state is set to low voltage, which is seen as grounding. Inactive nodes are set to a high resistance state, which is equivalent to disconnection. The storage instance vector combines the states of edge nodes. Vector bits are changed to construct current transmission paths corresponding to different storage instances, representing their connected paths.

\subsection{Formation of Connected Paths as Current Flows Through Nodes}
\begin{figure}[htbp]
    \centering
    \includegraphics[width=6cm]{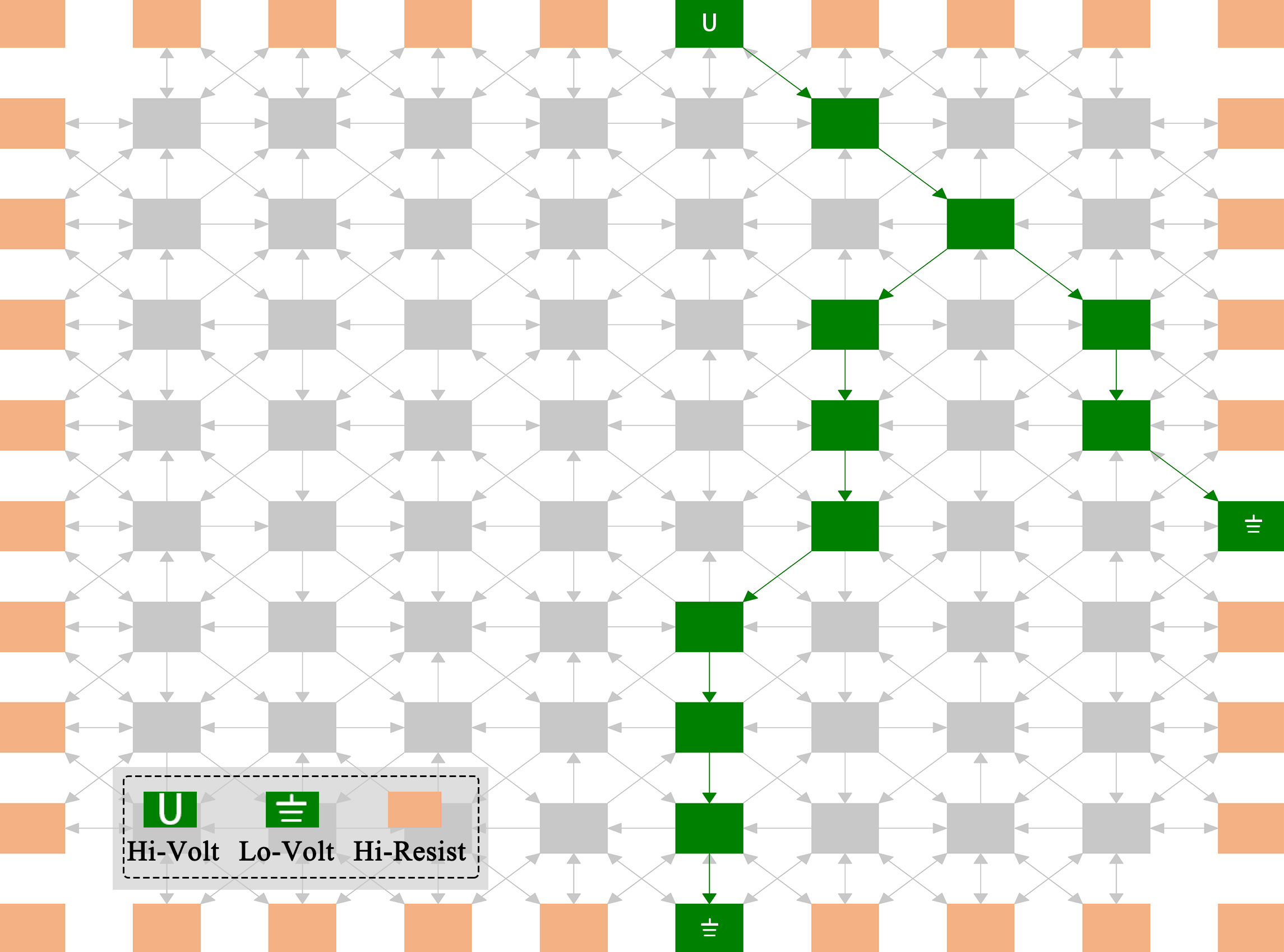}
    \caption{Setting Different Voltage States for Edge Nodes}
    \label{fig:setnode}
\end{figure}

In Figure \ref{fig:setnode}, some edge nodes in the directed graph are set to high voltage, some to low voltage, and the rest to a high resistance state, which represents the uploading of an instance vector onto the directed graph. Under the influence of the electric field, positive charges move from high to low potential, forming connected paths through internal nodes. Current flows from high-voltage nodes to low-voltage nodes. A storage instance is distributed in the directed graph in the form of current transmission paths. The green path in the graph has the fastest potential drop and is the dominant path of the storage instance. The other edge nodes remain in a high-resistance, non-conductive state.

\subsection{Node’s Adaptive Resource Competition Learning Algorithm}
According to the trace theory of memory, a neural network can reinforce recurring signal flow paths to achieve memory. After constructing the storage network through a directed graph and node circuit, converting storage instances into current transmission paths through potential differences, the storage function requires simulation of the process of consolidating the dominant path into the directed graph. This relies on the changing rules of variable resistors in each node during storage mode.

Variable resistors in storage mode follow changing rules.

The sum of the resistances of multiple variable resistors in a single node remains constant, indicating that the total resource amount is constrained, i.e.,
\begin{equation}
    \sum_{i=1}^{m} \sum_{j=1}^{n} R_{xi\_yj}=R_c
\end{equation}
where $R_c$ is a constant, and $m$ and $n$ are the respective numbers of input and output paths.

Like a rectangle, if the area on the left side increases, the area on the right side decreases. The resistance of variable resistors through which current greater than a threshold value $I_t$ flows gradually decreases. As the total resistance within a node is constant, the reduced resistance is evenly distributed to the other variable resistors in the same node,
\begin{equation}
\left\{
    \begin{array}{c}R_{xi\_yj}(t+1)=R_{xi\_yi}(t)+\Delta r_{xi\_yj}(t)\\
    \left\{
        \begin{array}{l} k_{xi\_yj}=1, \quad if\quad I_{xi\_yj}>i_t\\ 
        k_{xi\_yj}=0, \quad otherwise\\ 
        k=\sum_{i=1}^{m} \sum_{j=1}^{n} k_{xi\_yj} \\ 
        C=\sum_{i=1}^{m} \sum_{j=1}^{n} R_{xi\_yj}(t)\times k_{xi\_yj}\times L_r
        \end{array}
    \right.\\
    \Delta r_{xi\_yi}(t)=
    \left\{
        \begin{array}{l} -R_{xi\_yj}(t)\times L_r, \quad if\quad I_{xi\_yj}>I_t\\
        \frac{C}{m\times n-k} , \quad otherwise
        \end{array}
    \right.\\
    \sum_{i=1}^{m} \sum_{j=1}^{n} \Delta r_{xi\_yj}(t)=0
    \end{array}
\right.
\end{equation}
where $t$ is the number of iterations, $R_{xi\_yi}(t)$ is the resistance of the variable resistor between input node $xi$ and output node $yi$ at the $t^{th}$ iteration, $\Delta r_{xi\_yi}(t)$ is the resistance change at the $t^{th}$ iteration, $I_{xi\_yj}$ is the current between input node $xi$ and output node $yi$, $k_{xi\_yj}$ indicates whether $I_{xi\_yj}$ exceeds the threshold $I_t$, $k$ is the total number of variable resistors where current is greater than the threshold, $L_r$ is the proportion of resistance reduction in resistors with current flow after one iteration, $L_r$ can be considered the learning rate, and $C$ is the total value of the reduced resistance.

\subsection{Path Current Calculation Based on Kirchhoff’s Laws}
To evaluate the effectiveness of storage, it is necessary to calculate the current flowing through each path. For complex meshed circuits, Kirchhoff’s Voltage Law (KVL) and Kirchhoff’s Current Law (KCL) can be applied. KVL states that the algebraic sum of the potential differences (voltages) across all elements in a closed loop is zero, and KCL states that the sum of currents entering a node is the sum of currents leaving it.
\begin{figure}[htbp]
    \centering
    \includegraphics[width=6cm]{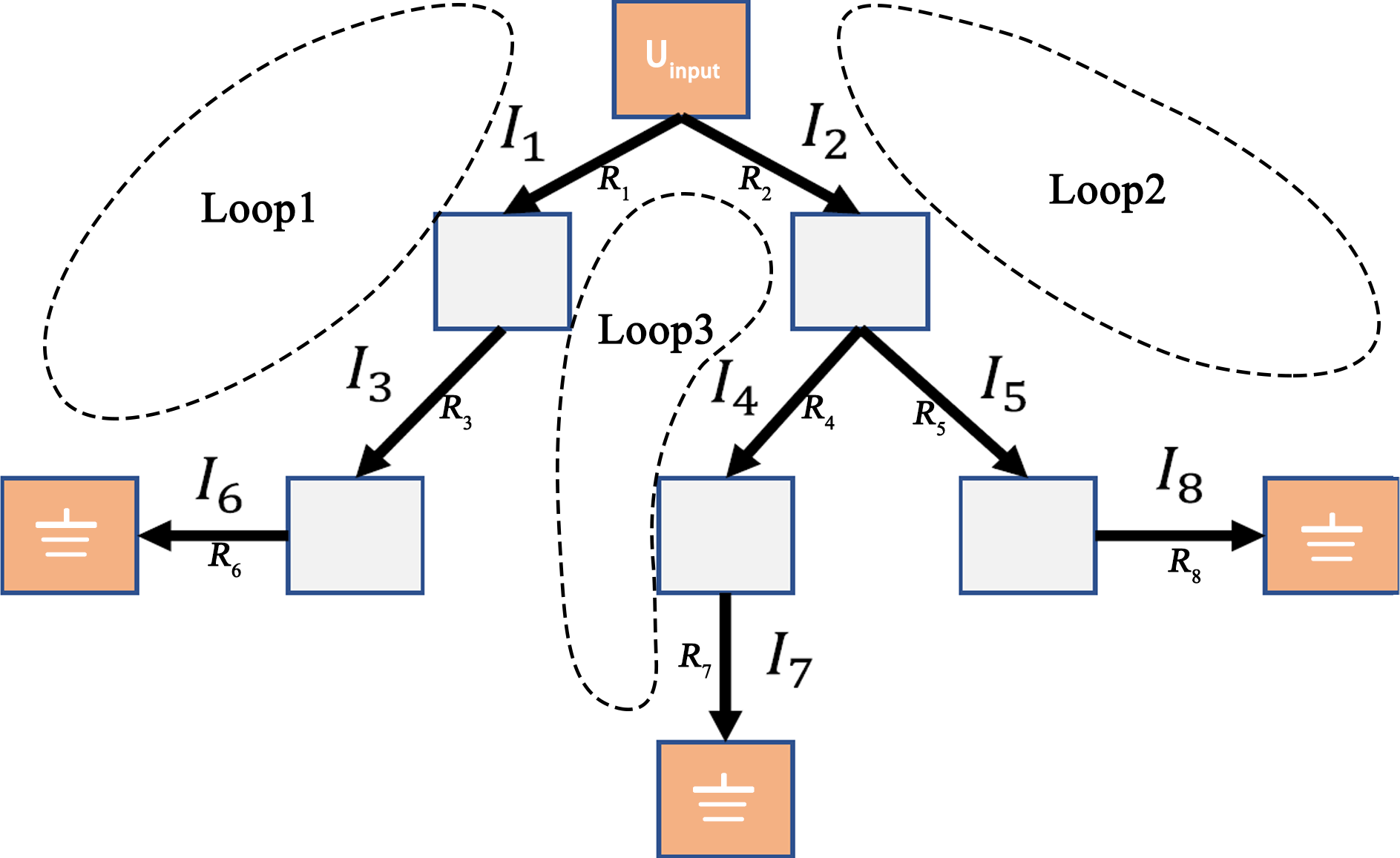}
    \caption{Currents on Multiple Connected Path Branches in Directed Graph}
    \label{fig:kvlkcl}
\end{figure}

In the directed graph, we know the variable resistors of each path and the input voltage. The positive pole of the input power source is connected to the nodes in the input state and the negative pole to the nodes in the output state. By applying KVL in the loop from the output-state edge nodes to the input-state edge nodes and using KCL at the internal nodes, a set of equations can be established to solve for the current values in each branch. Nodes that are not activated can be considered to have no current flow and can be ignored in the calculations. Taking the current flow paths in the directed graph in Figure \ref{fig:kvlkcl} as an example, the input and output currents of each node are numbered. The equations obtained by applying KCL to internal nodes are
\begin{equation}
\left\{
    \begin{array}{c}
    I_1-I_3=0\\
    I_2-I_4-I_5=0\\
    I_3-I_6=0\\
    I_4-I_7=0\\
    I_5-I_8=0\\
    \end{array}
\right.
\end{equation}
and, applying KVL to loops, we obtain
\begin{equation}
\left\{
    \begin{array}{c}
    I_1\times R_1+I_3\times R_3+I_6\times R_6-U_{input}=0\\
    I_2\times R_2+I_5\times R_5+I_8\times R_8-U_{input}=0\\
    I_2\times R_2+I_4\times R_4+I_7\times R_7-U_{input}=0\\
    \end{array}
\right.
\end{equation}
The voltage is calculated using Ohm’s Law, $U=IR$, where $U_{input}$ is the input voltage, and $R_i$ is the resistance value of the variable resistor on the corresponding path within the node for current $I_i$.

Through the above process, we can evaluate the retrieval effect of storage instances by calculating the current flowing through each path in retrieval mode. A significant current still flowing through the dominant path in retrieval mode indicates successful retrieval. We next discuss the storage and retrieval processes.

\subsection{Storage and Retrieval Process of Current Flowing Paths}
The storage and retrieval of paths simulate the respective biological processes of memory and recall. According to a node’s adaptive learning algorithm, in storage mode, the resistance of variable resistors through which a substantial current continually flows will gradually decrease. Continuously uploading the same storage instance vector to the edge nodes will establish stable current transmission paths between nodes from high to low voltage. Among the paths within each node, the path with the fastest potential drop will dominate, and the resistance of variable resistors on this dominant path will gradually decrease through natural competition, while the resistance on other paths increases. This change solidifies the current transmission path traces in the directed graph. The longer the current flow, the lower the resistance, making it easier for current to flow, for the path to be more likely activated, and for the stored content to be more profound, reinforcing the path.

In retrieval mode, some bits of the storage instance vector may be lost, such as remembering an object’s shape but not its color. When such a partial valid vector, the so-called probe vector, is loaded onto the directed graph, only certain edge nodes are set to their corresponding states, and lost edge nodes are set to low voltage. This assumes the possible existence of uncertain features, and later judgments are made based on the current level, with larger output currents indicating retrieved paths. The input state bits in the probe vector must not be lost; otherwise, no retrieval can occur due to the lack of input. In extreme cases, all bits except those corresponding to the input nodes in the probe vector may be lost. From Ohm’s Law, sub-paths with smaller resistance receive more current. Since the resistance on the dominant path of the storage instance is relatively low, a larger proportion of current will flow along this path. The longer the duration of the storage process, the less the resistance on the path, and the greater the proportion of current flowing along the dominant path, making selective current flow the basis for path retrieval.

\begin{figure}[htbp]
    \centering
    \includegraphics[width=6cm]{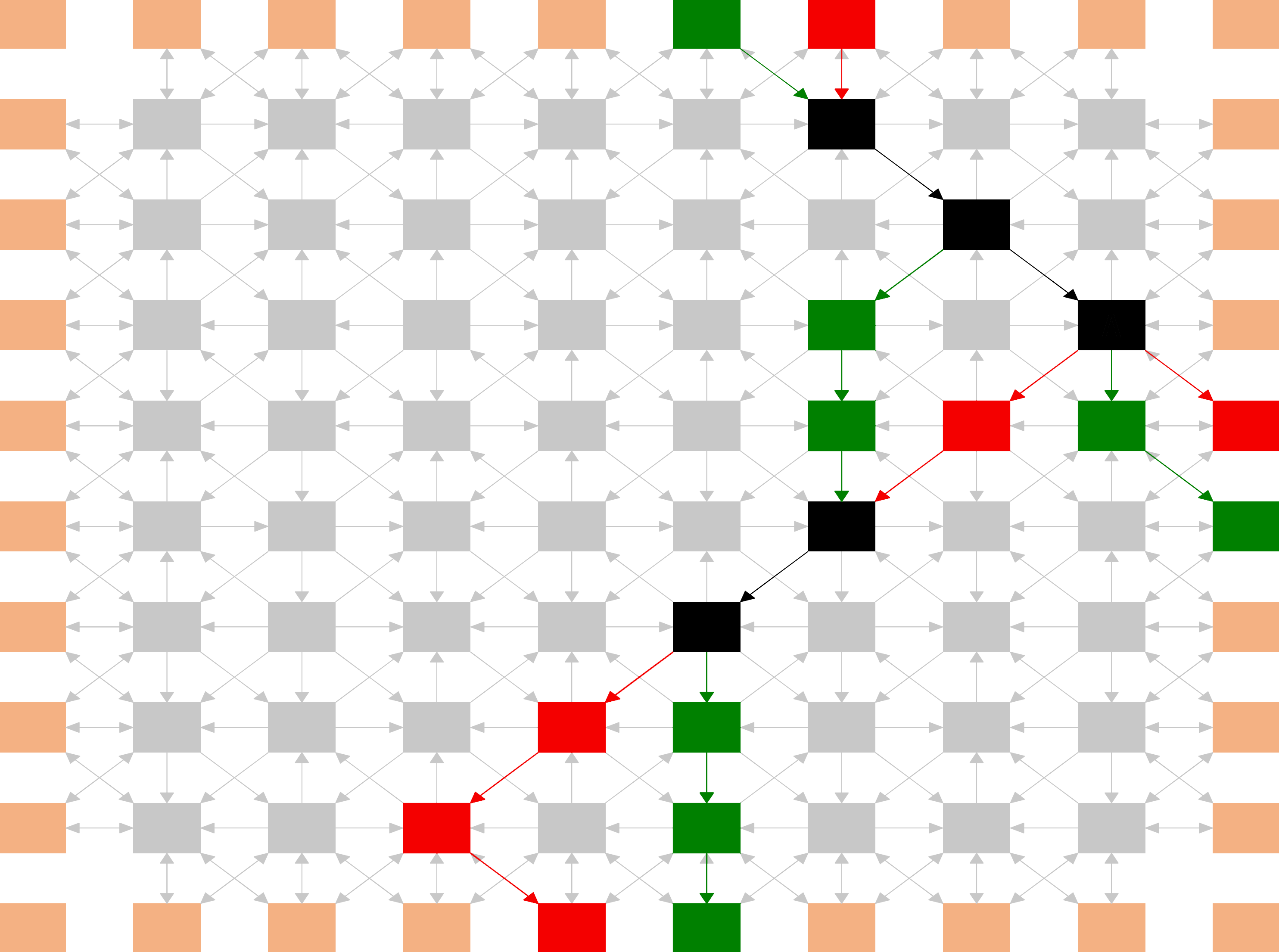}
    \caption{Distribution of Two Storage Instance Paths in Directed Graph}
    \label{fig:2memory}
\end{figure}

Notably, when more than one storage instance is stored in a directed graph and their current flow paths overlap, new stored instances will disrupt the path traces of the original instances solidified in the graph, in a process known as retroactive inhibition. In such cases, for accurate retrieval, the number of permissible lost bits in the probe vector decreases, which is like requiring more detailed features to distinguish similar objects. For example, from Figure \ref{fig:2memory}, it can be seen that, after sequentially uploading green and red instances into the directed graph, since the paths overlap (black path), the graph stores the composite paths of the two storage instances.

\section{Simulation Experiment}
We conducted a series of simulation experiments to investigate the role of directed graphs in modeling memory systems. Our focus was on exploring the generation, functionality, and capacity of these graphs, simulating the varied microstructural complexities found in biological memory systems. The experiments ranged from initial graph generation to comprehensive storage functionality tests, including capacity testing for single and multiple memory instances. We also conducted comparative analyses with other models.
\subsection{Generation of Directed Graph}
Considering that the memory systems of biological entities have many microscopic structural differences, for better simulation, we first initialize the directed graph, to generate directed graph networks of varying scales and connection relationships. For ease of visualization, the directed graph is set in a grid format, with its width and depth determining its size and scale. After determining the number of nodes based on the side length, the node matrix is traversed, and connections in random directions are established between adjacent nodes to generate directed graph networks with different topological structures, as shown in Algorithm \ref{alg:1}. Figure \ref{fig:base10} shows an example of randomly generated directed graph networks with a side length of 10.

\begin{algorithm}[ht]
\SetAlgoLined
\KwResult{Directed graph node matrix represented by adjacency list}
\textbf{Input:} Width and Depth of Directed Graph Node Matrix \(x, y\)\;
\textbf{Output:} Directed graph node matrix\;
\SetKwFunction{FMain}{genGraphNodeMatrix}
\SetKwProg{Fn}{Function}{:}{}
\Fn{\FMain{\(x,y\)}}{
    Initialization: Node matrix nodeMatrix with width and depth \(x, y\). For each node in the matrix, attributes include input node list inList, output node list outList, bidirectional node list 2wList\;
    \For{\(i \leftarrow 1\) \KwTo \(x\)}{
        \For{\(j \leftarrow 1\) \KwTo \(y\)}{
            The current node is represented by selfNode\;
            neighbors \(\leftarrow\) getNeighbor(selfNode)\;
            neiLen \(\leftarrow\) len(neighbors)\;
            \For{neighbor \(\leftarrow\) neighbors[0] \KwTo neighbors[neiLen - 1]}{
                \(xN, yN\) represent the horizontal and vertical coordinates of the node neighbor in the matrix\;
                \If{the edge between node selfNode and neighbor has not been assigned a direction}{
                    \eIf{not (isEdge(selfNode) and isEdge(neighbor))}{
                        direct \(\leftarrow\) Random Boolean value\;
                        \eIf{direct}{
                            nodeMatrix[i][j]['inList'].add(neighbor)\;
                            nodeMatrix[xN][yN]['outList'].add(selfNode)\;
                        }{
                            nodeMatrix[i][j]['outList'].add(neighbor)\;
                            nodeMatrix[xN][yN]['inList'].add(selfNode)\;
                        }
                    }
                    {
                        nodeMatrix[i][j]['2wList'].add(neighbor)\;
                        nodeMatrix[xN][yN]['2wList'].add(selfNode)\;
                    }
                }
            }
        }
    }
    \KwRet{nodeMatrix}\;
}
\textbf{End Function}\;
\caption{Directed Graph Generation Algorithm}
\label{alg:1}
\end{algorithm}

\subsection{Assigning Activation States to Edge Nodes}
\begin{figure}[htbp]
    \centering
    \includegraphics[width=6cm]{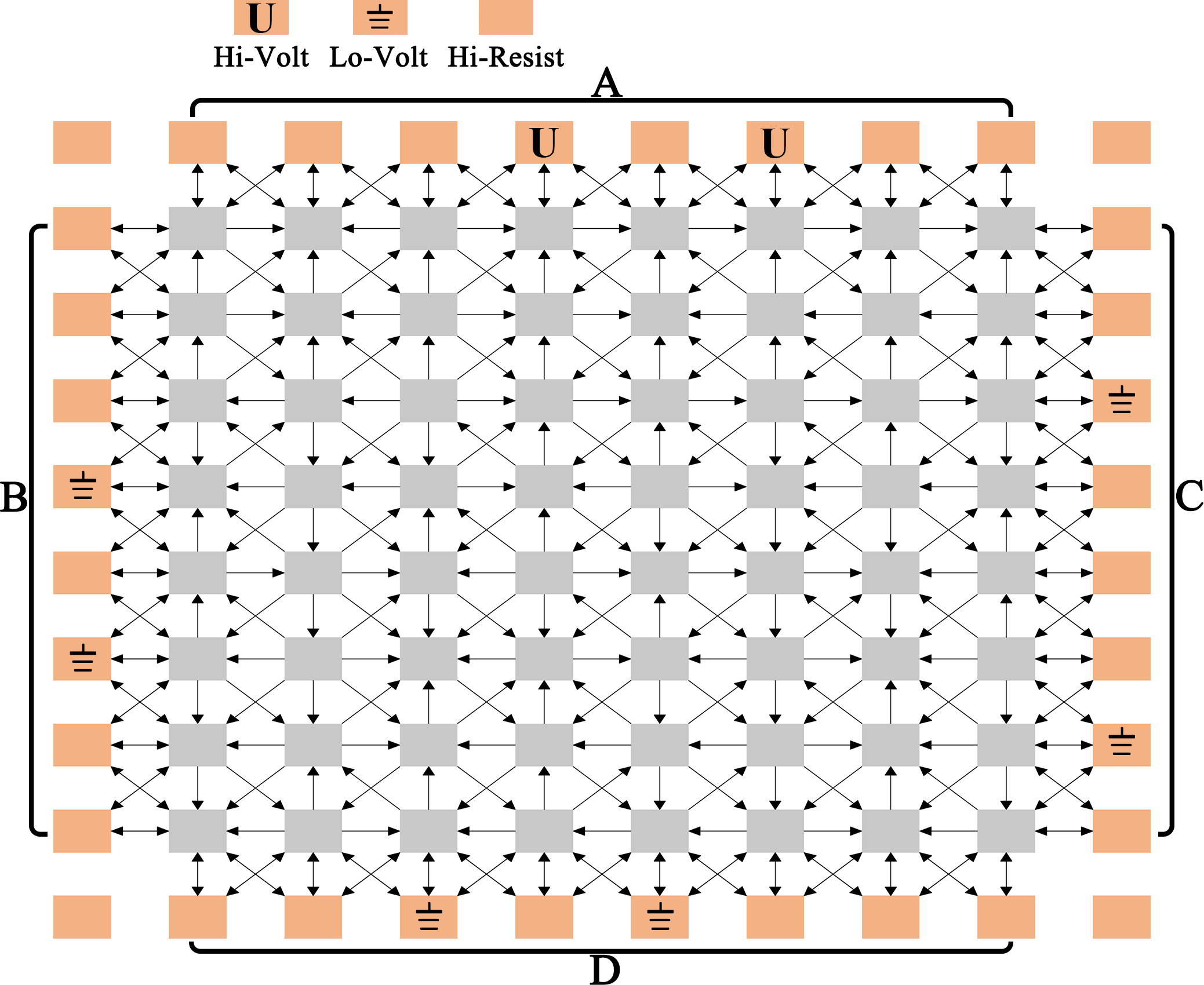}
    \caption{Correspondence of Storage Instance Vector to Edges of Directed Graph}
    \label{fig:v2edge}
\end{figure}

In biological neural networks, different stimuli lead to the formation of memories. In our simulation, a storage instance is uploaded through the edge nodes of the directed graph. For visualization purposes, the top edge of the directed graph is defined as the input edge, where activated nodes are assigned a high-voltage state. The others are output edges, with activated nodes assigned a grounding state. Nodes corresponding to nonzero vector bits are activated. Inactive edge nodes are assigned a high-resistance state. Each storage instance is divided into component A, which corresponds to the input edge, and components B, C, and D, corresponding to the three output edges. In the directed graph network, electric current flows from A to B, C, and D. Like water, the current always flows downstream, where A acts as the upstream, and B, C, D as the downstream. Figure \ref{fig:v2edge} shows a schematic representation of the correspondence between each component of a storage instance and the directed graph.

In storage mode, stable current can only form on the paths between nodes in a high-voltage state and grounding state. According to their changing rules, as the current continues to flow, the resistance values of variable resistors on the dominant path with the fastest potential drop decrease continuously. The path is solidified into the directed graph through the reduction of resistance values of variable resistors on the path.

Retrieval has two modes: awakening based on upstream information, and awakening based on both upstream and downstream information. However, in either mode, the edge node corresponding to component A, serving as the energy source of the directed graph, must not be omitted from the probe vector. The difference between the probe vector and the storage instance vector is quantified by the Hamming distance, $d=\rho n$, where $n$ is the length of the vector, and $\rho$ is the proportion of differing bits.

When using awakening based on upstream information, all bits in the probe vector, except for the input bit corresponding to component A, are set to be missing, testing the effect of retrieval in the most extreme case. In this mode, the mapping method of the edge nodes corresponding to component A remains unchanged, and all edge nodes corresponding to the output state of B, C, and D are grounded to simulate the complete loss of this characteristic information, intending to activate all output paths. Here, $\rho$ reaches its maximum, $\rho_{max}=1-\frac{x+f\times(2y+x)}{n}$, where $x$, $y$ are the respective width and depth of the directed graph, and $f$ is the proportion of activated nodes in the storage instance.

When multiple instances are stored in the directed graph, there is interference between the paths of each instance, including partial sharing, partial destruction, or even breakage. The original paths are no longer complete, and continuing to use awakening based on upstream information may lead to retrieval failure. Therefore, awakening based on both upstream and downstream information is introduced. Compared with awakening based on upstream information, the probe vector in this mode retains more information and has fewer missing bits, i.e., some edge nodes corresponding to the output state of components B, C, and D remain in their original high-resistance state. This aims to reduce available output paths, better retrieve the composite paths stored in the directed graph, and improve the retrieval effect of each instance path.

\subsection{Storage Functionality Testing}
We tested the storage functionality of the directed graph when storing a single instance. A storage instance was loaded into the directed graph network to observe if it could be successfully stored. The experiment checked if the current on the connected path of the storage instance in the directed graph during upstream information-based awakening was significantly greater than that of other paths. This allowed for a clear distinction between old and new paths. For a directed graph with only one storage instance, if the storage is successful, it should be able to accurately restore the original path. Since the resistance values on the original path are relatively small, only the previously reinforced dominant path of the storage instance will maintain a larger proportion of current, while the current on other paths will be smaller. This is an important criterion to differentiate between established and temporary traces.

\begin{figure}[htbp]
    \centering
    \includegraphics[width=6cm]{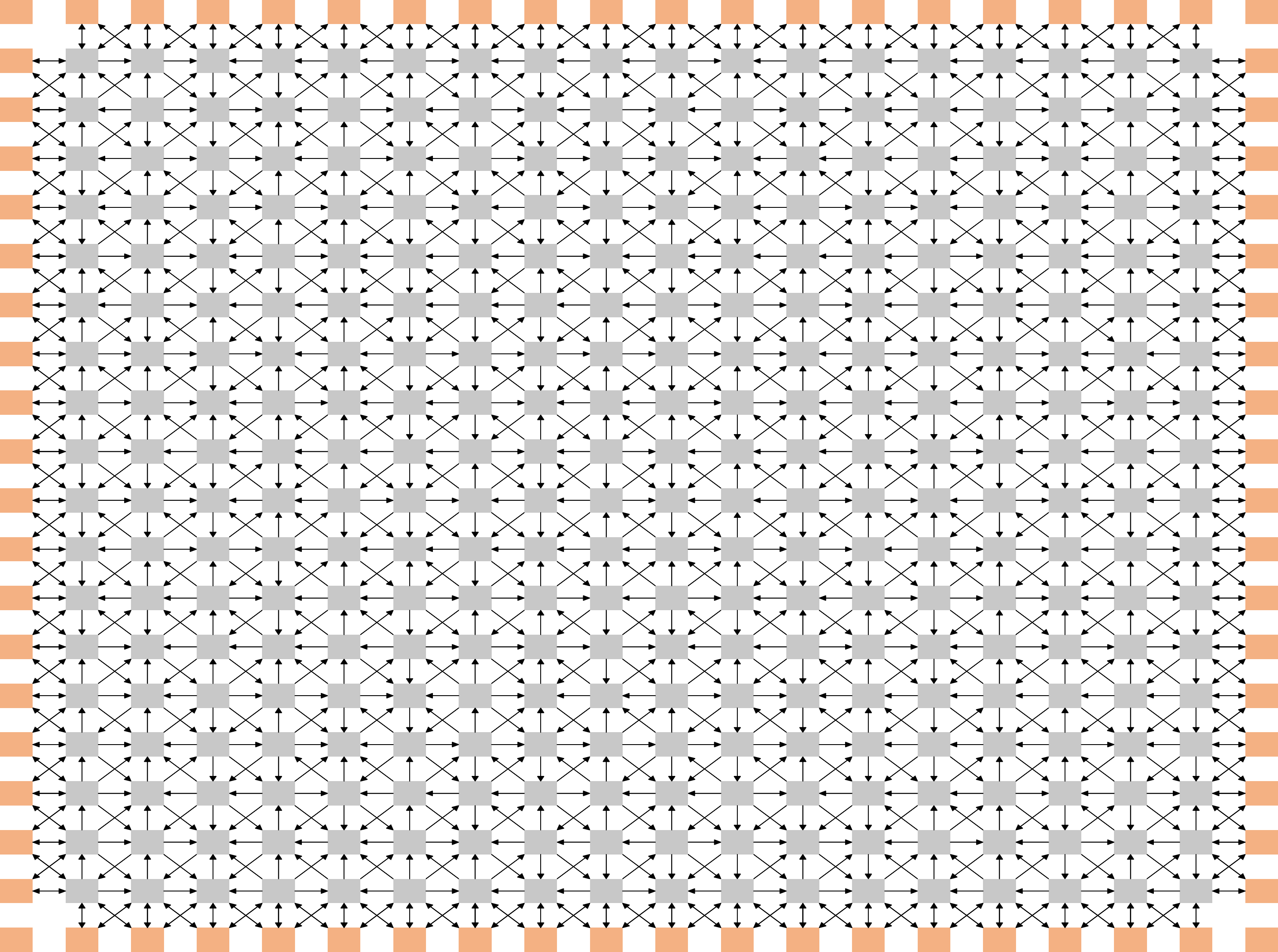}
    \caption{Randomly Generated 20 $\times$ 20 Directed Graph Where Directed Connections Between Nodes Form the Basis for Path Creation}
    \label{fig:base20}
\end{figure} 

In this experiment, the learning rate $L_r$ was set at 10\%. The storage functionality was tested in a 20 $\times$ 20 directed graph network, as shown in Figure \ref{fig:base20}, which was randomly generated using Algorithm \ref{alg:1}. In each test, a storage instance, with 30\% of the edge nodes randomly activated, was uploaded to the directed graph, for 30 iterations of the storage process, like rehearsing to reinforce memory. This learning process iteratively adjusts the resistance values of the variable resistors on connected paths, gradually forming a so-called dominant path with relatively low resistance. After each iteration, the graph was switched to retrieval mode, the probe vector was uploaded, and the output current of each path was calculated. The proportion of the current of the storage instance’s dominant path to the total current of all paths is
$$\alpha=\frac{\sum_{i=1}^{k}i_j}{\sum_{j=1}^{m}i_j},m>k$$
where $i_j$ represents the output current values of each path, $m$ is the total number of paths, and the dominant path consists of the first $k$ paths. At this point, $\rho$=0.525.

\begin{figure}[htbp]
    \centering
    \subfigure[Proportion of Current Output on Dominant Path of Storage Instance]
        {\includegraphics[width=6.5cm]{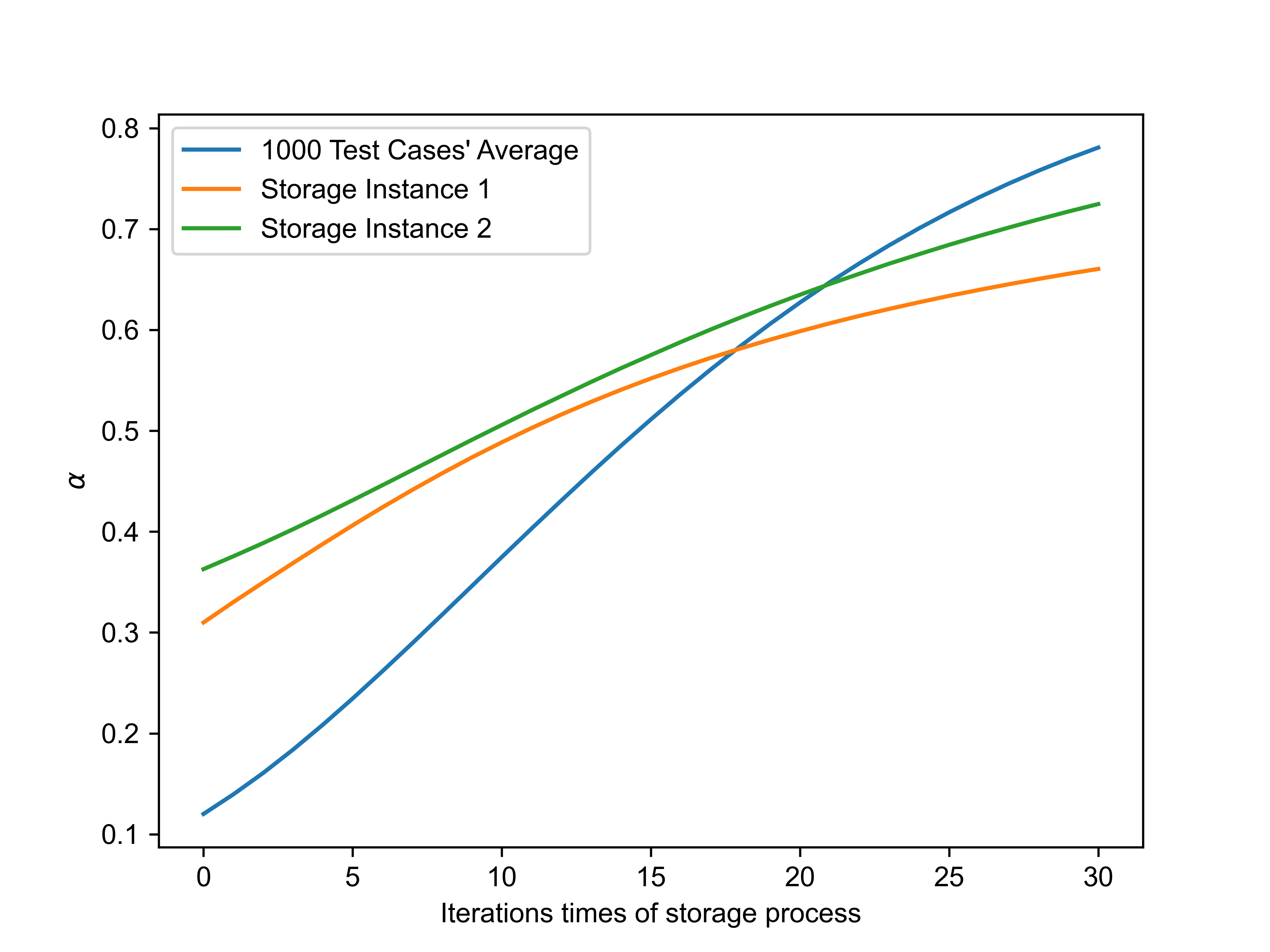}\label{fig:currrntprop}}
    \hspace{5mm}
    \subfigure[Changes in Variable Resistors within Single Node]
        {\includegraphics[width=6.5cm]{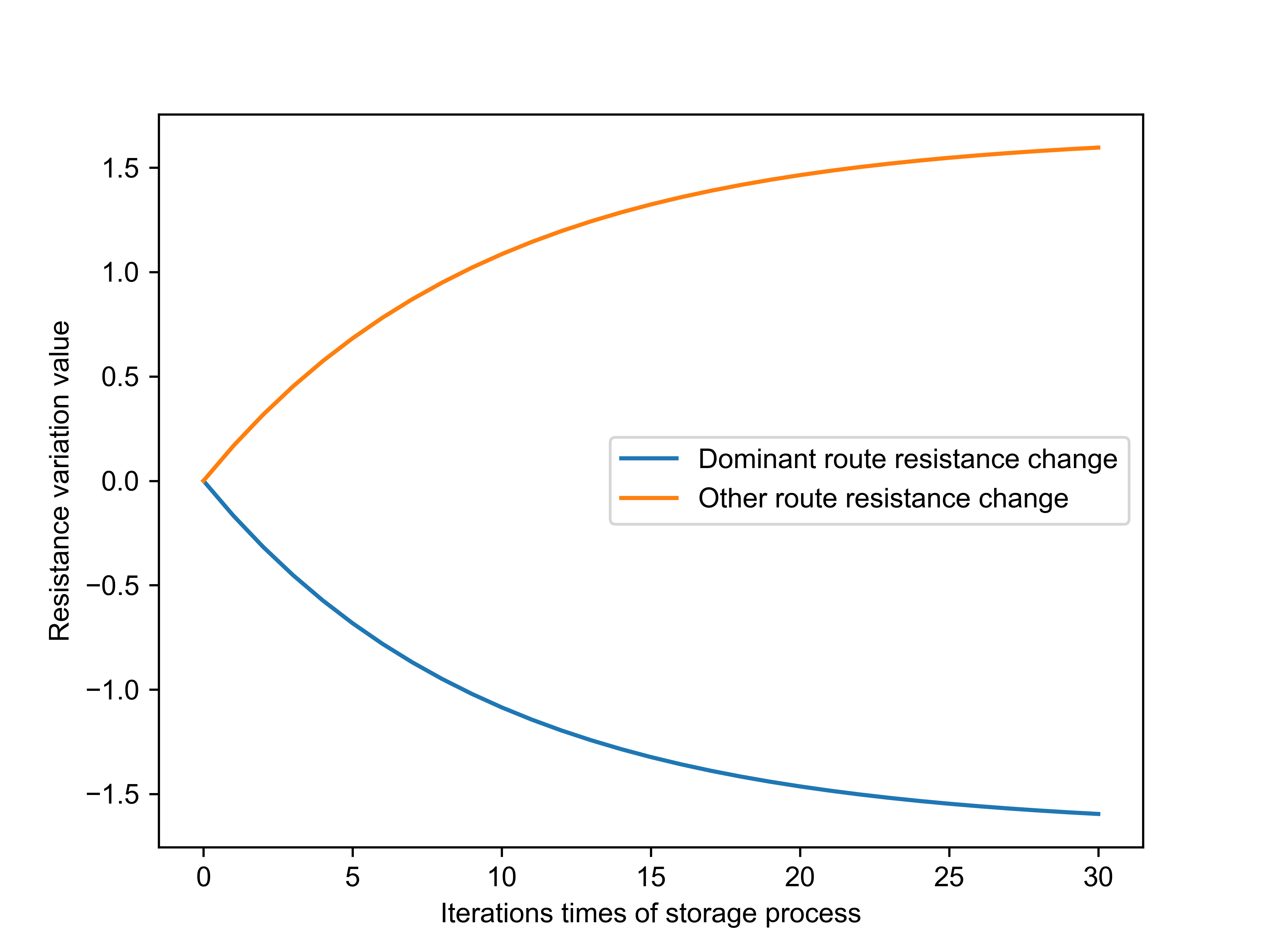}\label{fig:reschange}}
    \caption{Changes in Directed Graph Metrics at Different Iteration Counts}
    \label{fig:iterchange}
\end{figure}

\begin{figure}[htbp]
    \centering
    \subfigure[Instance 1]{
        \includegraphics[width=6cm]{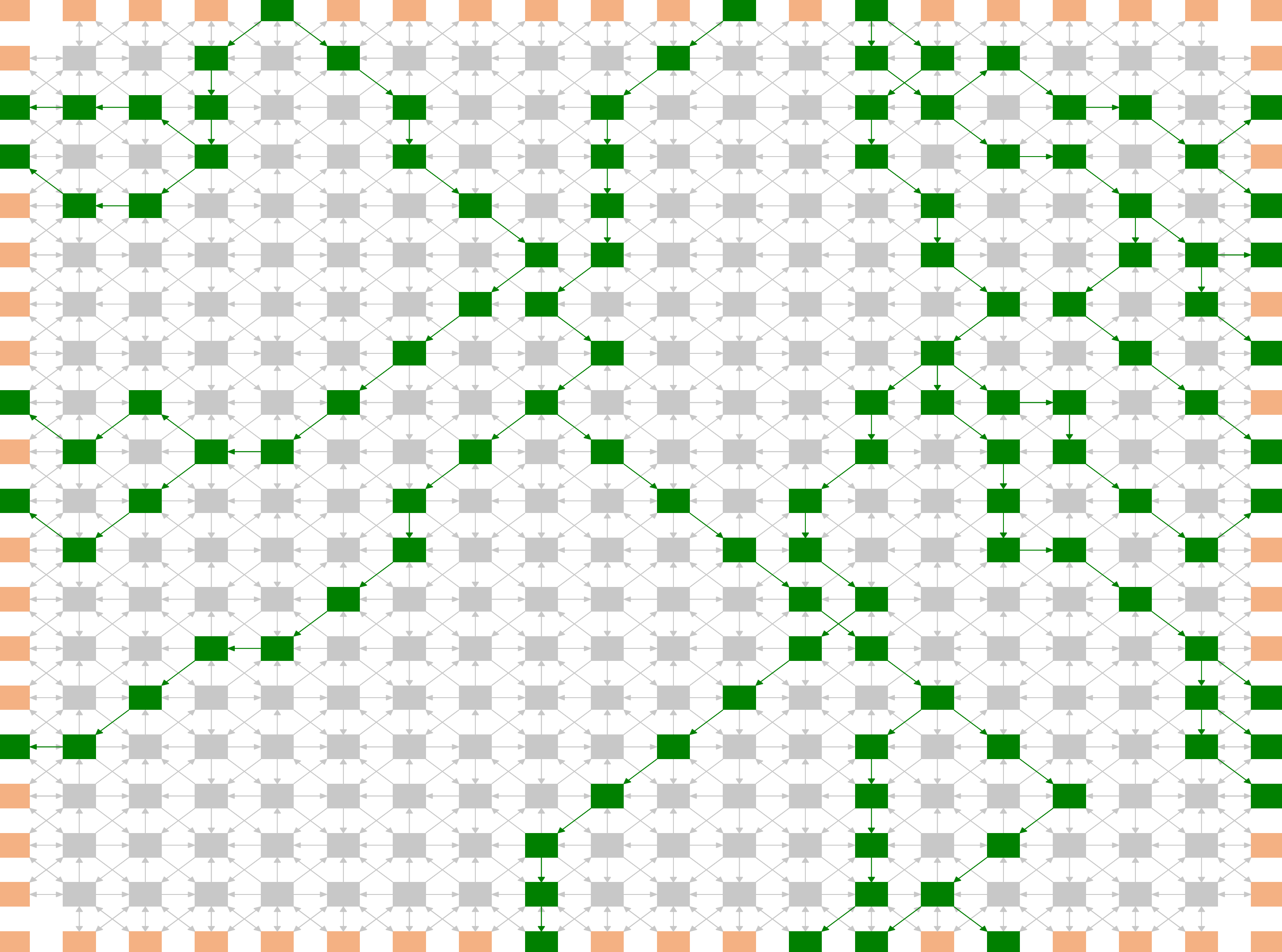}
    \label{fig:ins1}
    }
    \hspace{5mm}
    \subfigure[Instance 2]{
        \includegraphics[width=6cm]{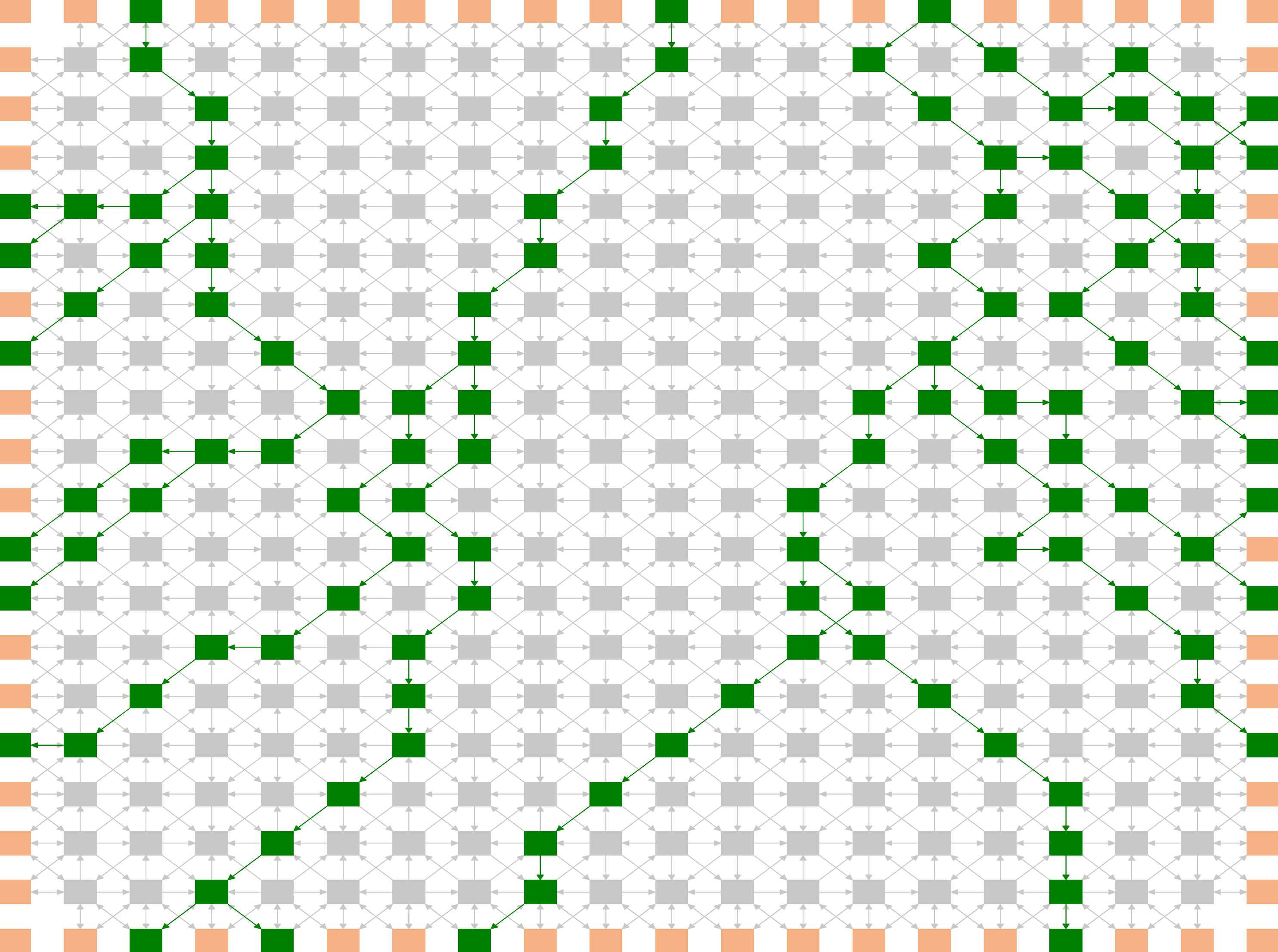}
    \label{fig:ins2}
    }
    \caption{Dominant Path of Storage Instance in Directed Graph}
\end{figure}

The results, as shown in Figure \ref{fig:currrntprop}, indicate that with increasing iterations, randomly generated storage instances 1 and 2 have differences in the $\alpha$ curve after iterative reinforcement, but both achieve significant improvement, similar to the varying efficiency humans experience when memorizing different content. A random node was selected to observe the changes in the resistance values of its internal variable resistors, as depicted in Figure \ref{fig:reschange}. The resistance values of variable resistors on the dominant path inside the node gradually decreased, and the reduced resistance was evenly distributed to the variable resistors on the other paths of the same node, causing the resistance on these paths to increase gradually. The dominant paths of storage instances 1 and 2 in the directed graph are shown in Figures \ref{fig:ins1} and \ref{fig:ins2}. Figure \ref{fig:currrntprop} also shows the average value of $\alpha$ from experiments with 1000 test cases, which showed an increase after iteration. This demonstrates that the storage function of the directed graph operates well. With different storage instances, after a certain number of iterations, the current flowing through the original dominant path in retrieval mode is significantly greater than that of other paths, solidifying this path as the representation of the storage instance, thereby realizing the storage function. The experimental results indicate that storage instances 1 and 2 require a similar number of iterations to achieve the same storage effect. Random testing on another 1000 examples yielded similar results, indicating that the directed graph and path-finding algorithms are not dependent on specific examples and are universally applicable to a wide variety of content.

\begin{figure}[ht]
    \centering
    \subfigure[$0^{th}$ Iteration]{
        \includegraphics[width=4cm]{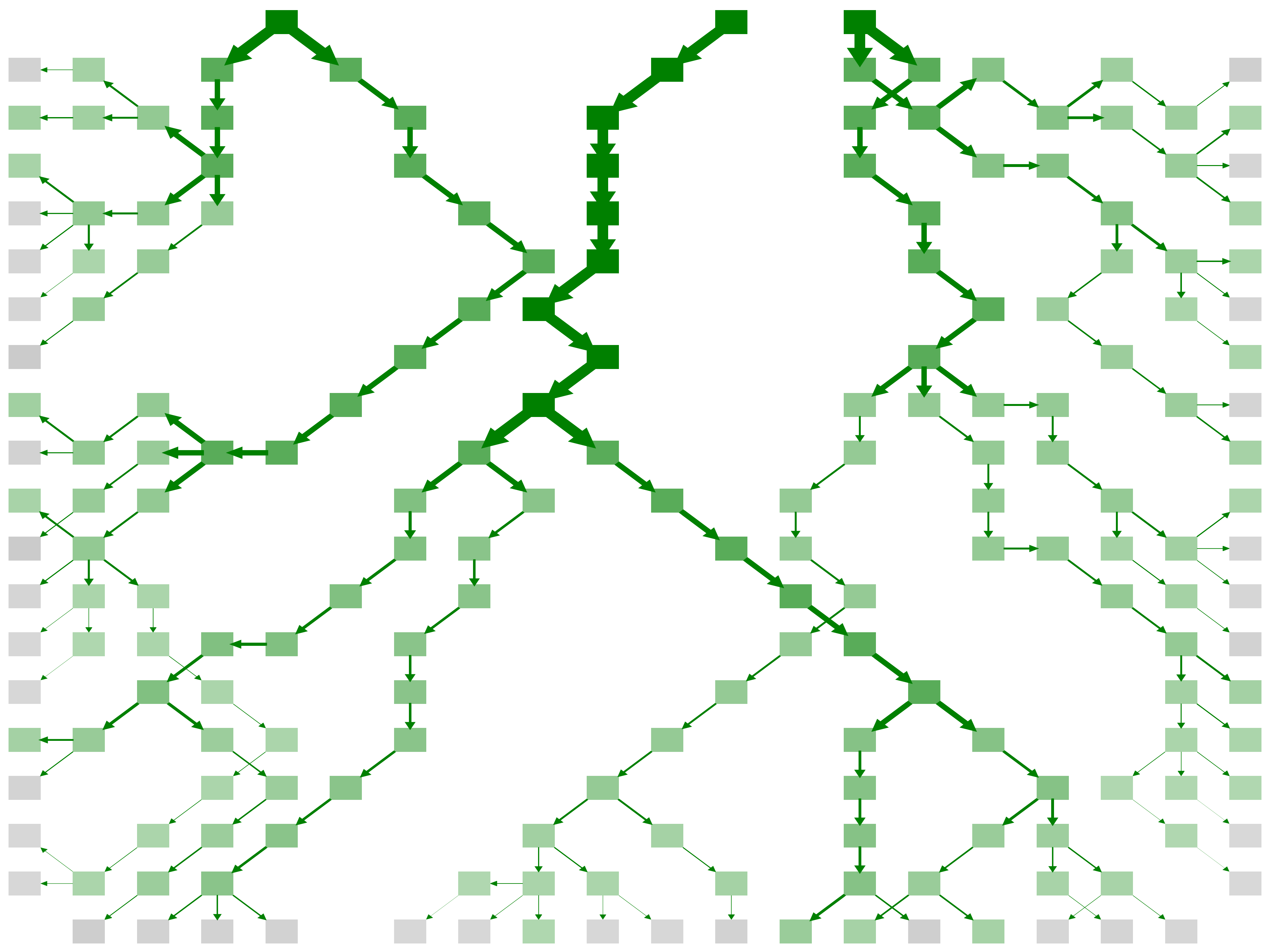}
    }
    \subfigure[$10^{th}$ Iteration]{
        \includegraphics[width=4cm]{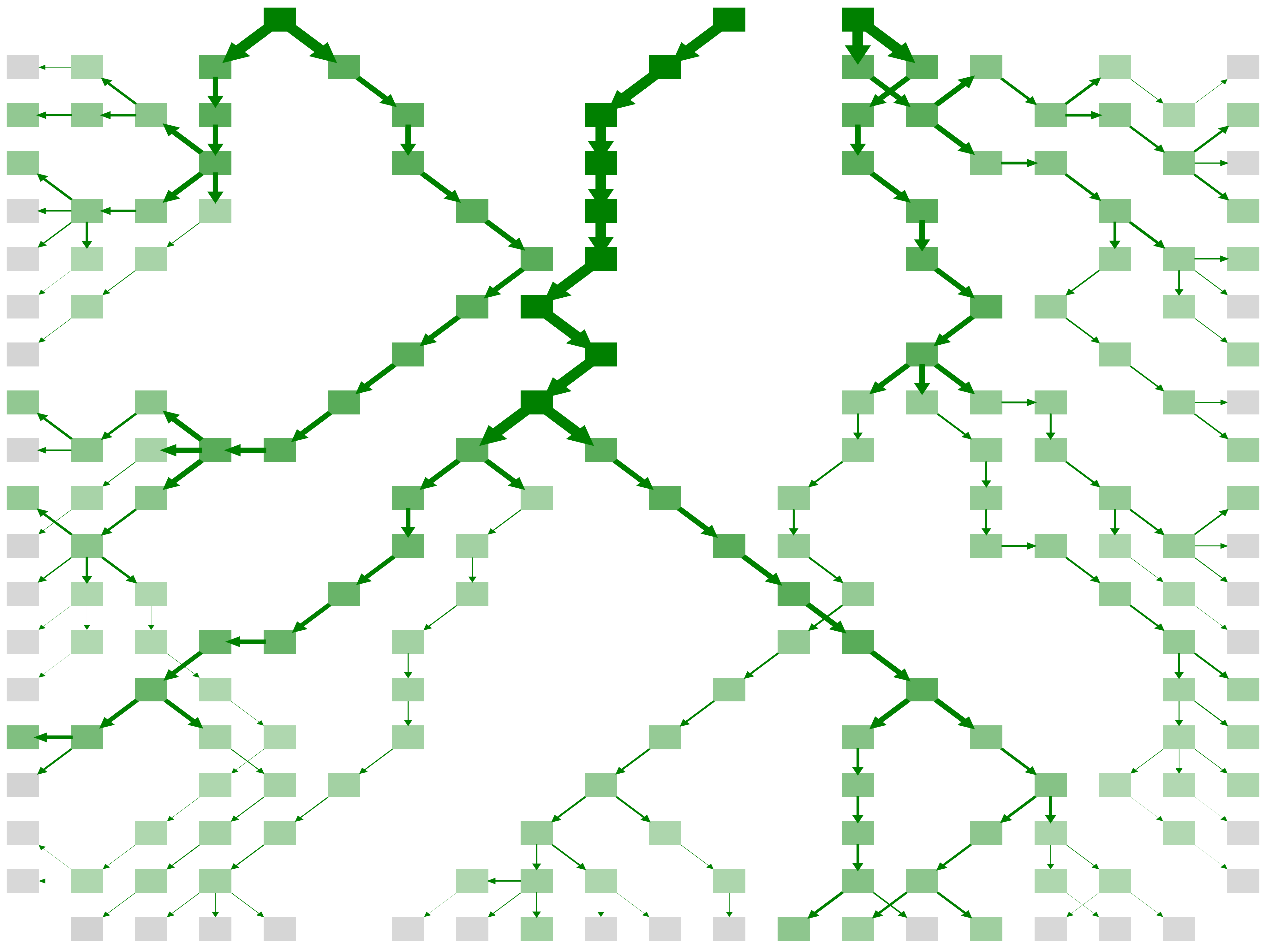}
    }
    \subfigure[$20^{th}$ Iteration]{
        \includegraphics[width=4cm]{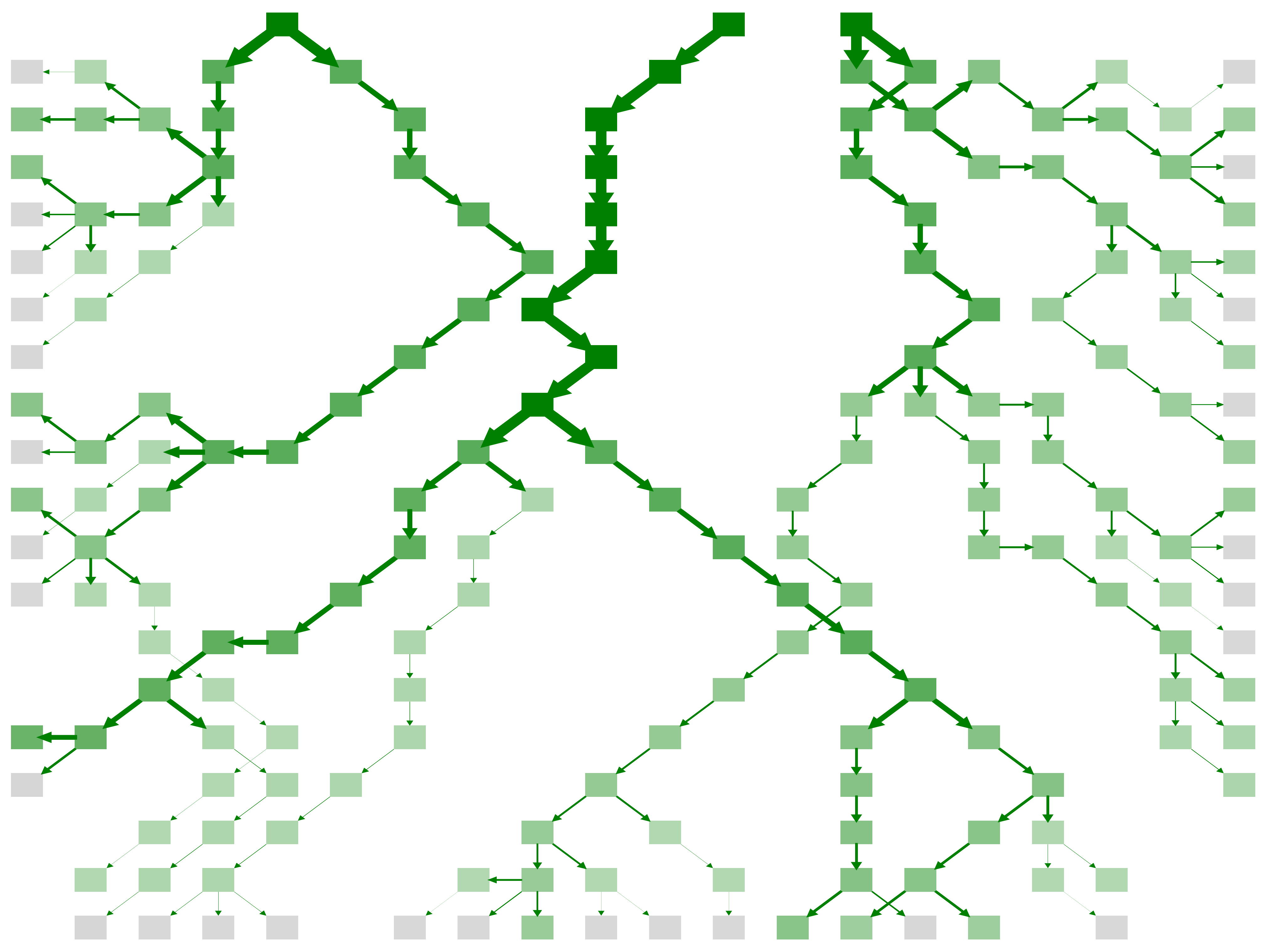}
    }
    \subfigure[$30^{th}$ Iteration]{
        \includegraphics[width=4cm]{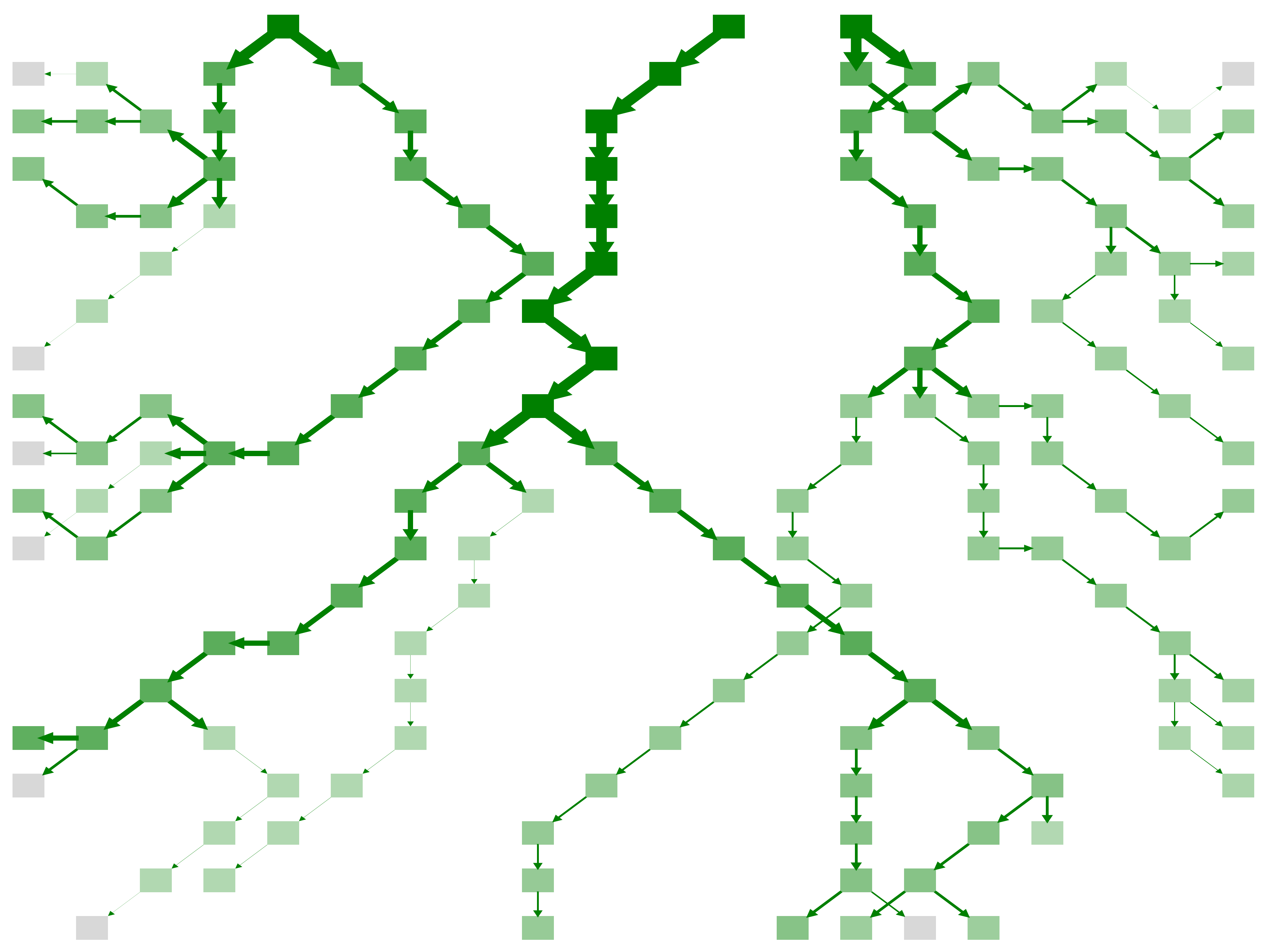}
    }
    \caption{Currents of Various Paths at Different Numbers of Iterations}
    \label{fig:currvisure}
\end{figure}

Taking storage instance 2 as an example, the currents in various paths were calculated and visualized at the 0th, 10th, 20th, and 30th iterations of storage, visually demonstrating the process by which the dominant path gradually becomes more prominent among many paths with increasing iterations, as shown in Figure \ref{fig:currvisure}, where the thickness of paths and opacity of nodes represent the magnitude of the current. Thicker paths and lower opacity indicate greater current. Green paths have the fastest potential drop from the activated input nodes to each connected output node. Paths with a current less than 0.4 mA are not displayed. Green edge nodes are activated output nodes corresponding to the storage instance, and gray nodes are inactivated output nodes. It can be observed that as iterations proceed, the current on the dominant path continuously increases, and the path is gradually solidified into the graph.

\begin{figure}[htbp]
    \centering
    \subfigure[]{
    \includegraphics[width=6cm]{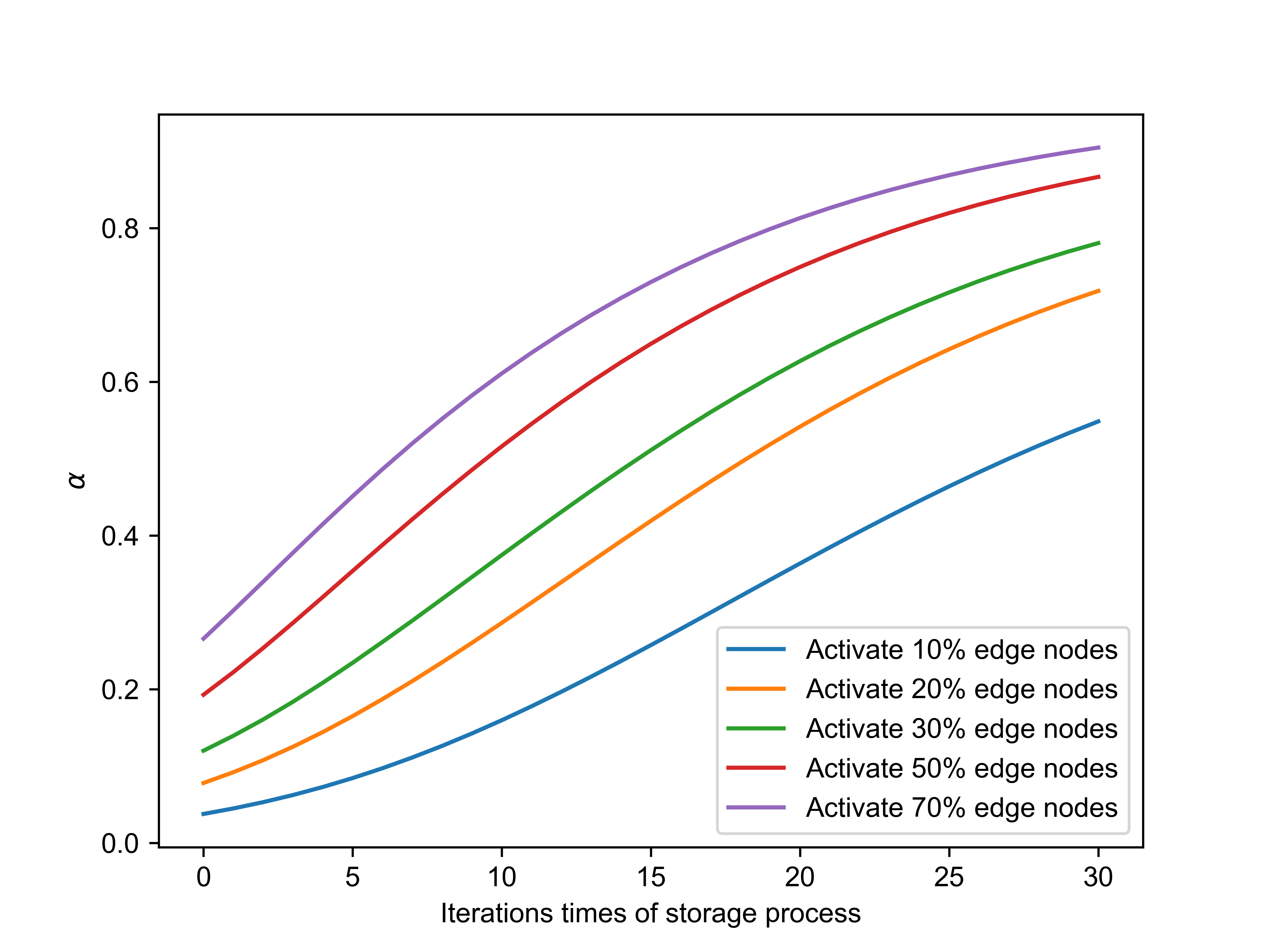}
    \label{fig:actratio} 
    }
    \hspace{5mm}
    \subfigure[]{
    \includegraphics[width=6cm]{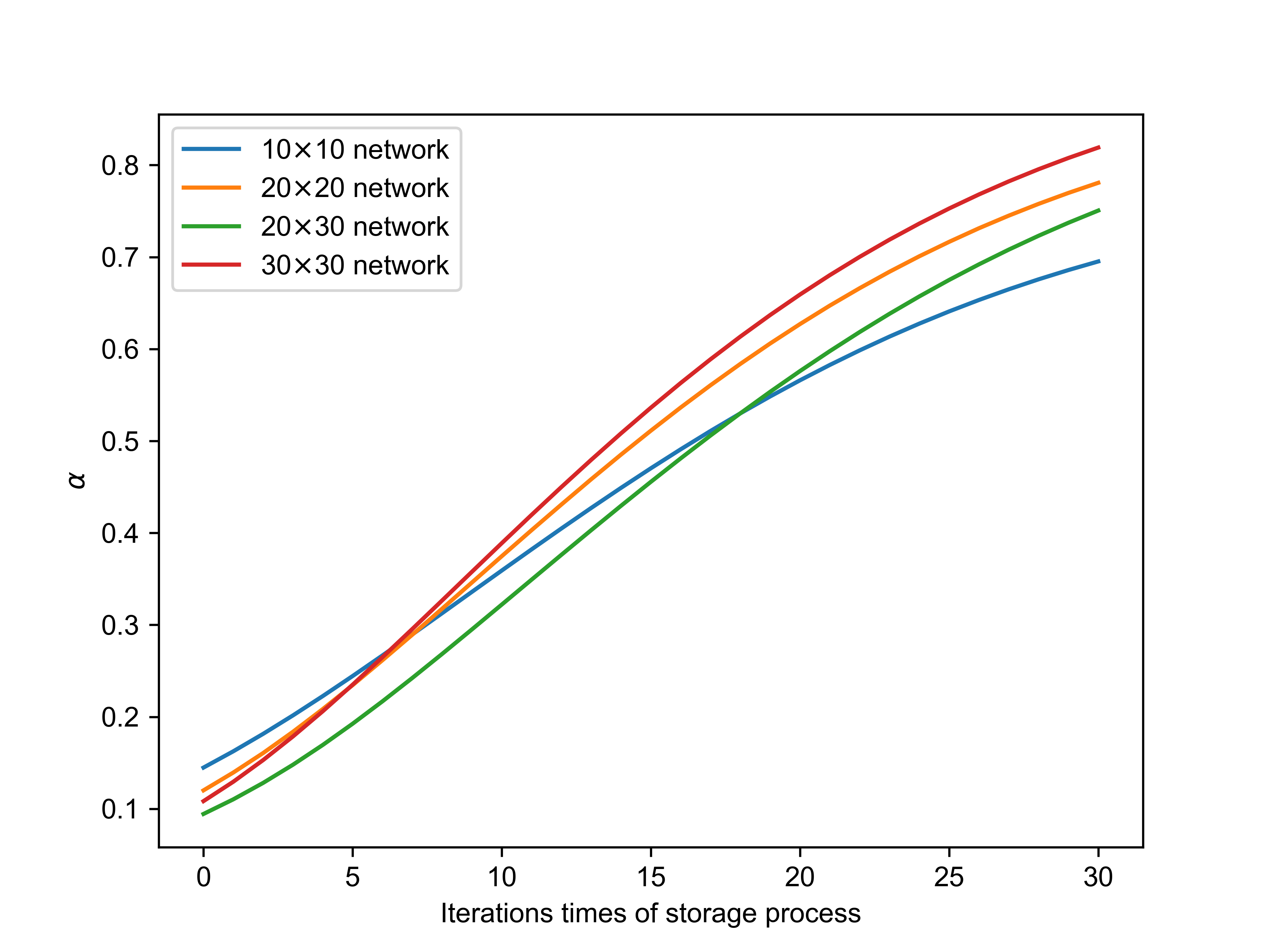}
    \label{fig:universal}
    }
    \caption{(a):Proportion of Current Output on Dominant Paths Under Different Node Activation Ratios (b):Directed Graphs of Different Sizes and Topological Connections Demonstrate Similar Learning Capabilities, Indicating Universality}
\end{figure}

The activation proportion of edge nodes affects the number of directed graph resources occupied by a storage instance. A higher activation proportion may lead to the occupation of more paths, and the dominant path may be more expansive. We conducted experiments on storage instances with different activation proportions of edge nodes. The experimental results are the average value of $\alpha$ with 1000 test cases, as shown in Figure \ref{fig:actratio}, indicating that storage instances with higher proportions have higher proportions of current output $\alpha$ on their dominant paths after storage. This may be because such instances can access more path resources for output during retrieval. At the same time, according to the definition of $\rho_{max}$, the higher the proportion of activated nodes, the smaller $\rho_{max}$ becomes, reducing the difference between the probe vector and the original vector.

\begin{figure}[htbp]
    \centering
    \subfigure[30 $\times$ 30]{
        \includegraphics[height=5cm]{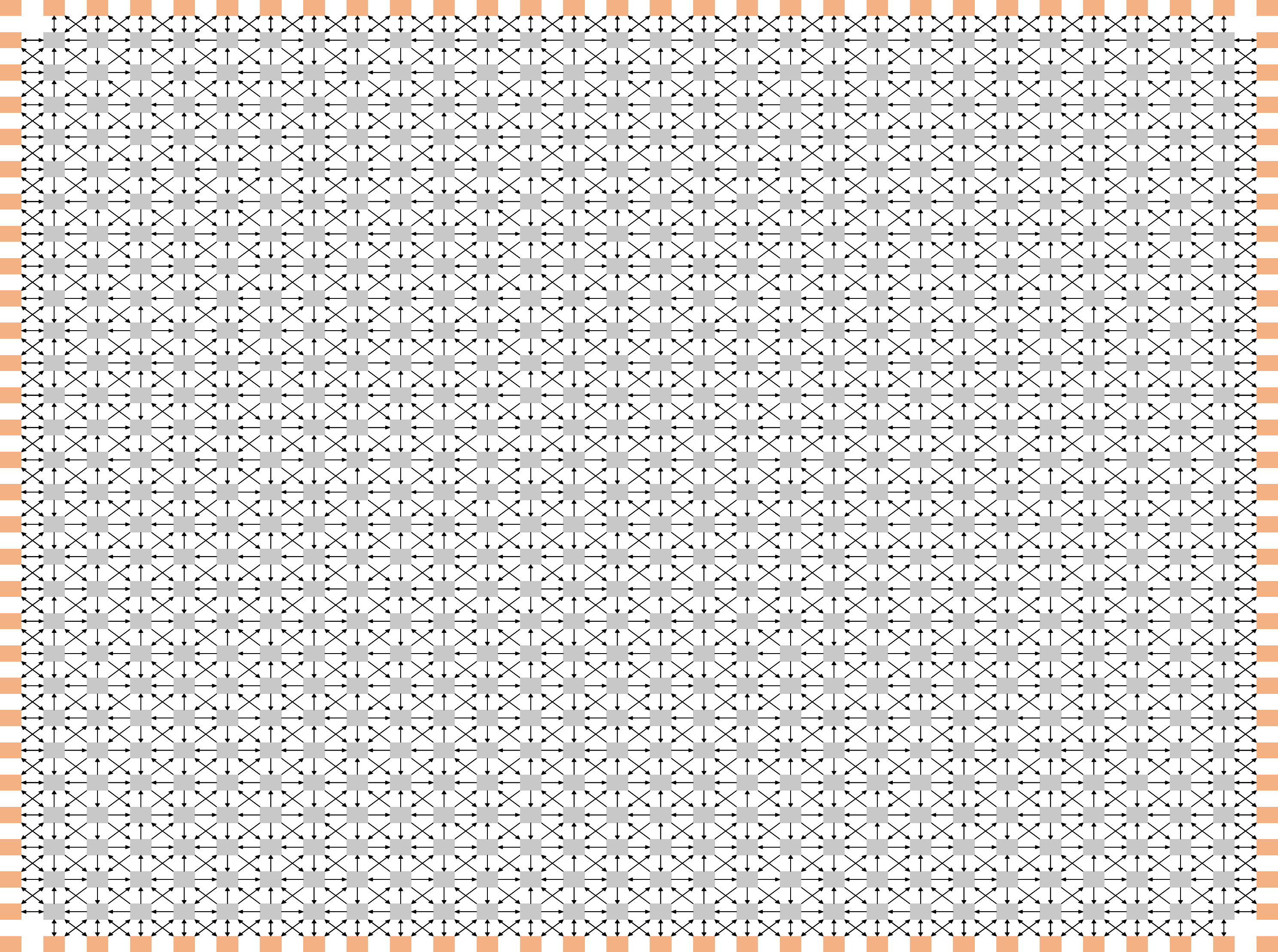}
    \label{fig:base30}
    }
    \hspace{5mm}
    \subfigure[20 $\times$ 30]{
        \includegraphics[height=5cm]{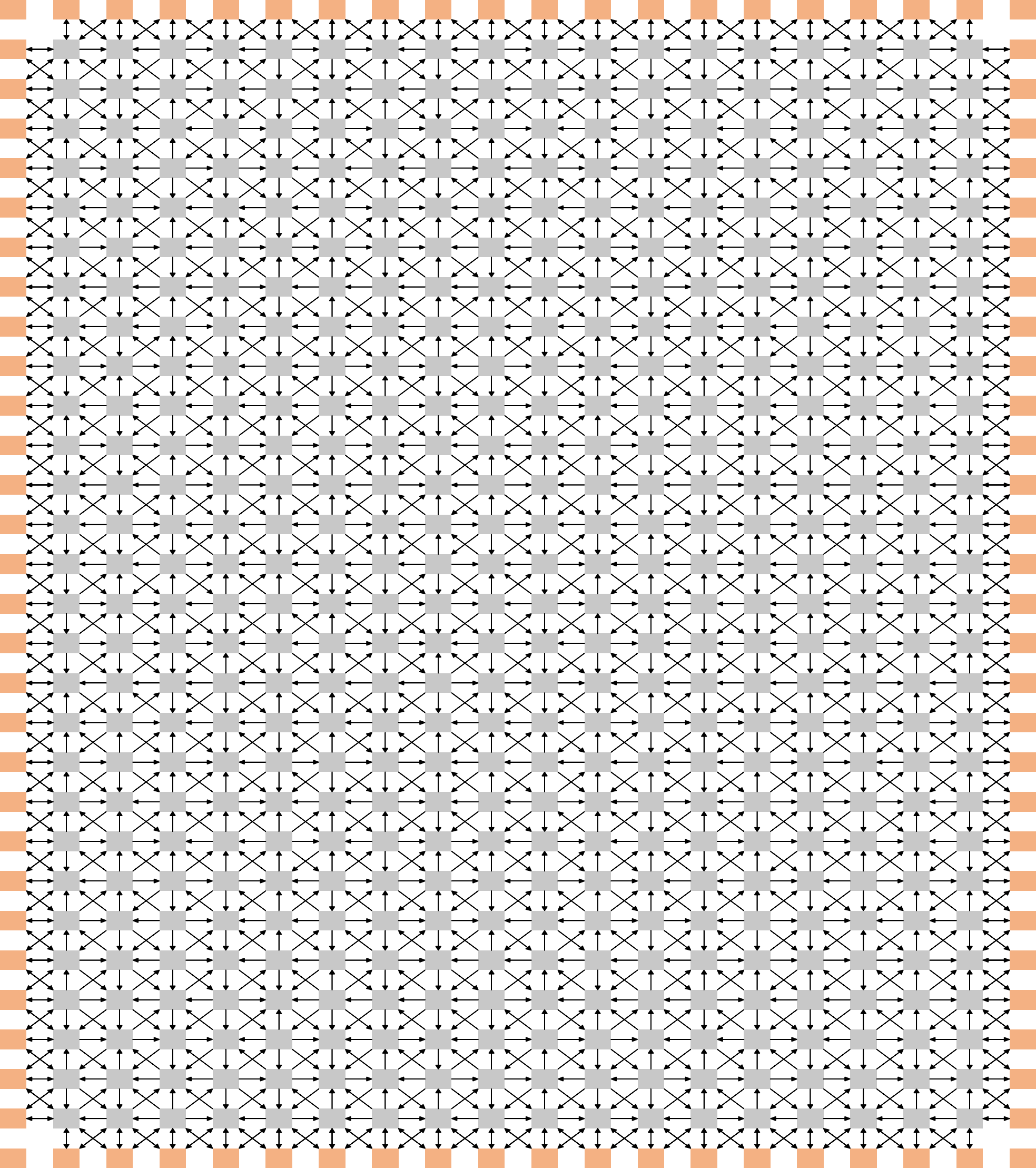}
    \label{fig:base2030}
    }
    \caption{Directed Graphs Randomly Generated with Significant Variations in the Number of Nodes and Connectivity Relationships}
\end{figure}

Neural networks in different people’s brains vary in terms of the number of neurons, topological connections, and other microscopic details. Despite these numerous and fine structural differences, they do not seem to affect memory function, indicating universality. To verify the effectiveness of the storage function of the directed graph network under different topological structures and sizes, Algorithm \ref{alg:1} was used to generate randomly topologized directed graphs of sizes 10 $\times$ 10, 30 $\times$ 30, and 20 $\times$ 30, as illustrated in Figures \ref{fig:base10}, \ref{fig:base30}, and \ref{fig:base2030}, respectively. Using the parameters from Figure \ref{fig:iterchange}, storage experiments were conducted in these directed graphs, and the average $\alpha$ was calculated from 1000 test cases. The results, as shown in Figure \ref{fig:universal}, indicate that despite the different sizes and topologies of the three directed graphs, they all achieve improvements after iterations, meaning they all can perform the storage function.

The results show that as the number of iterations in the storage process increases, the current flowing through the dominant path of the storage instance gradually increases. This indicates that the path is progressively solidified into the directed graph after the sample is repeatedly uploaded. The learning algorithm may exhibit slight differences in storage effectiveness in directed graph networks of different sizes and topological structures, but it can generally achieve memory of path traces. This demonstrates that directed graphs generally possess storage functionality when combined with path representation and path solidification algorithms.

\subsection{Incremental Storage and Retrieval Testing of Multiple Storage Instances}
Both the human brain and computer storage devices contain a large amount of varied content, and their storage mechanisms must ensure compatibility. In computers, file management systems ensure that different files exclusively occupy different storage spaces, making compatibility issues relatively simple. However, in directed graphs, there is an issue of sharing small units. When more than one memory instance is stored in the same directed graph, the dominant paths of different memory instances will likely partially overlap. Subsequent stored memory instances will interfere with and potentially disrupt paths already stored due to resource competition, a phenomenon known as retroactive inhibition.

We conducted a two-part experiment in a 20 $\times$ 20 directed graph, as shown in Figure \ref{fig:base20}. First, testing was done using awakening based on upstream information, followed by tests using probe vectors with different $\rho$ values for awakening based on both upstream and downstream information. Storage instances with 30\% of edge nodes randomly activated were generated and incrementally stored in the same directed graph in a certain order. The learning rate $L_r$ was set at 10\%, and each instance was iterated 30 times in the storage process.

In the retrieval of each instance, the proportion of the output current of each branch on the dominant path to the total output current on the dominant path is calculated as
$$\beta_o=\frac{i_o}{\sum_{j=1}^{k}i_j},1\leq o\leq k$$
The primary evaluation metric was $\alpha$, while secondary metrics included the similarity of $\beta_1-\beta_k$ at this point to its value when the instance was stored alone in the same directed graph for the same number of iterations. Similarity was quantified using cosine similarity; the values of $\beta_1-\beta_k$ were arranged in the same order to form vectors, and the cosine similarity between them was calculated. This was used to assess the deviation in the output balance of each branch. These $\beta$ values form a vector,
$$V=[\beta_1\quad \beta_2\quad \cdots \quad \beta_k]$$
The cosine similarity is
$$\varphi = \frac{V_{pre} \cdot V_{curr}}{\|V_{pre}\|\|V_{curr}\|}$$
 where $V_{pre}$ and $V_{curr}$ refer to the vectors of $\beta$ values from the previous (stored alone) and current (with other instances stored) retrieval scenarios, respectively.
\begin{figure}[ht]
    \centering
    \subfigure[]{
        \includegraphics[width=5.5cm]{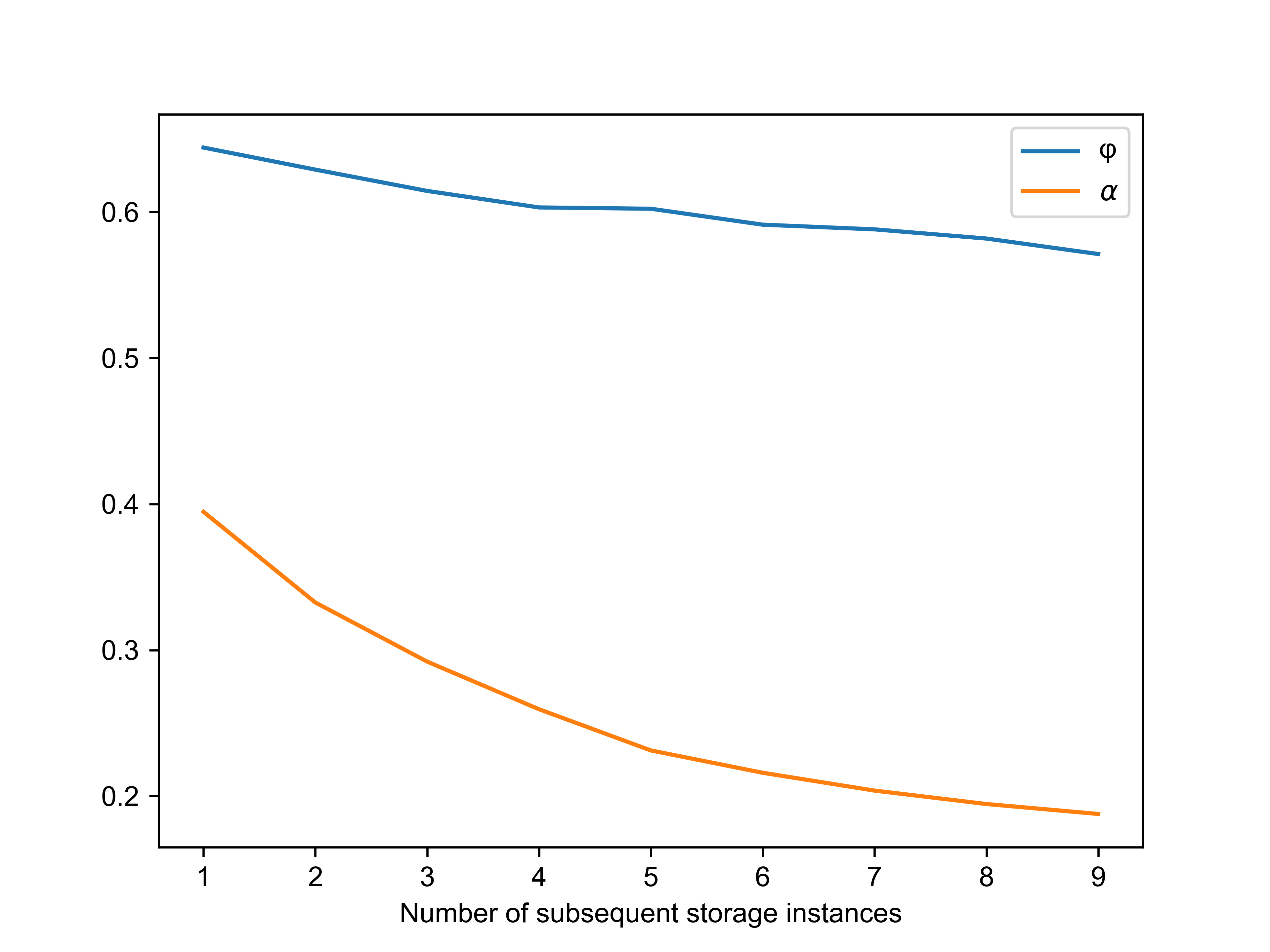}
        \label{fig:subsequenteffect}
    }
    \subfigure[]{
        \includegraphics[width=5.5cm]{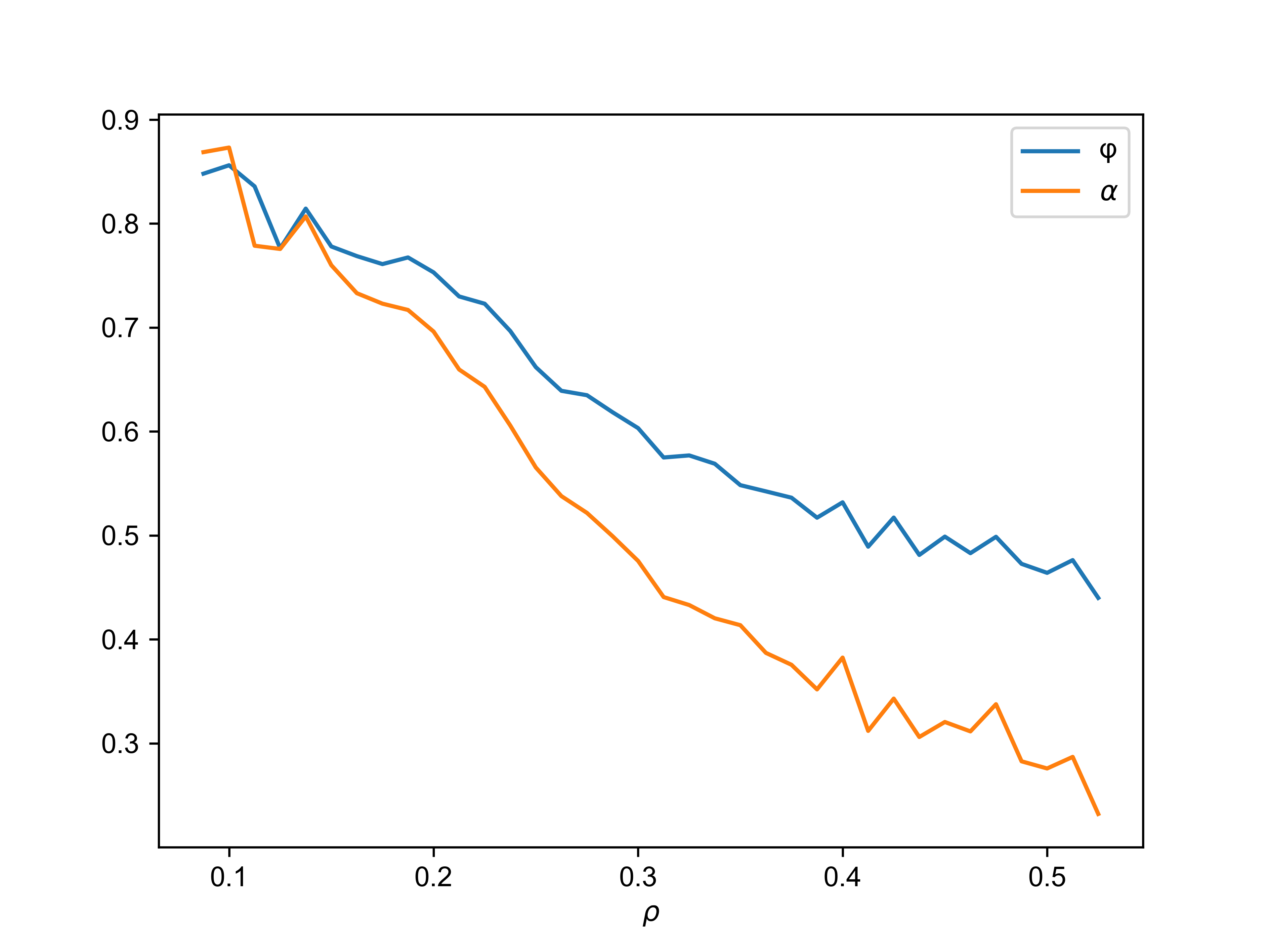}
        \label{fig:probeeffect}
    }
    \subfigure[]{
        \includegraphics[width=5.5cm]{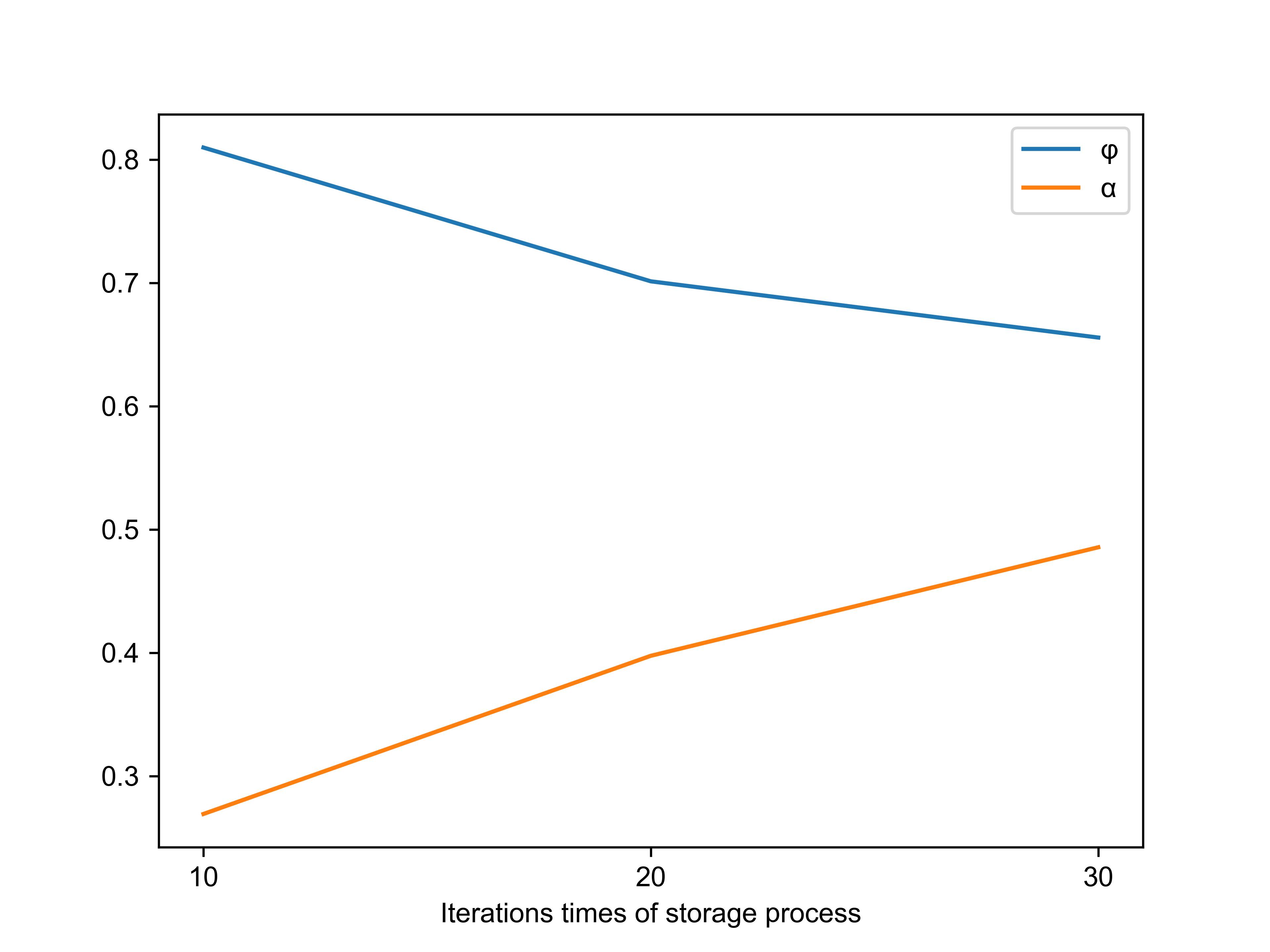}
        \label{fig:itereffect}
    }
    \caption{(a):Retrieval Effectiveness with Different Subsequent Storage Instance Counts (b):Impact of Probe Vectors with Different $\rho$ Values on Retrieval Effectiveness (c):Different Iteration Counts in Storage Process Influence Retrieval Effects}
\end{figure}

In the same directed graph, 10 memory instances were incrementally stored, and after completing the storage process of each instance, retrieval based on upstream information was conducted for the stored instances. The experiment was repeated 1000 times with different test cases, and the results were averaged. As shown in Figure \ref{fig:subsequenteffect}, the retrieval effectiveness significantly decreased compared with when stored alone, as discussed earlier. The number of subsequent instances (reflecting the original position of the extracted instance) showed a negative correlation with both $\alpha$ and $\beta$, indicating that the storage of new instances continuously interferes with paths already stored in the graph, but the impact on $\varphi$ is relatively minor.

The results of the experiment in Figure \ref{fig:subsequenteffect} also indicate that the directed graph retains the ability to distinguish whether a particular instance has been stored before, even after storing 10 instances. During retrieval, $\alpha$ remains higher than the average value shown in Figure \ref{fig:currrntprop} without iterative storage, indicating that the current still preferentially flows through the previously established paths. This is akin to recognizing a person whose name we cannot recall.

Testing based on both upstream and downstream information was conducted after all 10 memory instances were stored in each experiment. Probe vectors with different $\rho$ values were uploaded to test the retrieval effectiveness. The experiment was repeated 1000 times with different test cases, and the results were averaged. As shown in Figure \ref{fig:probeeffect}, the retrieval effectiveness gradually improves as $\rho$ decreases, with both metrics showing significant improvement. The results are markedly better than those from awakening based on upstream information alone. This is like a human better recalling things after receiving some hints.

From the previous experiments and the definition of the storage learning algorithm, it is known that the number of storage iterations and the learning rate affect the proportion of the dominant path output during retrieval. In an incremental storage process, the number of iterations for subsequent instances can also be considered an indicator of the degree of interference of preceding instances on solidified paths. To test the impact of the number of storage process iterations on the retrieval effectiveness of multiple storage instances, two storage instances with 30\% of edge nodes activated were randomly generated and uploaded to a 20 $\times$ 20 directed graph network, with the learning rate $L_r$ maintained at 10\%. The number of iterations in the storage process was set to 10, 20, and 30. The first and second instances stored were the disturbed and disturbing instance, respectively. After storage, the disturbed instance was subjected to upstream information-based awakening retrieval testing. The experiment was repeated 1000 times under different iteration counts, and the results were averaged. The results, as shown in Figure \ref{fig:itereffect}, indicate that the number of iterations in the storage process positively correlates with $\alpha$, and negatively with $\varphi$, which is consistent with theoretical analysis. The increase in $\alpha$ is due to the deeper solidification of the paths of the preceding instance, while the decrease in $\varphi$ is due to increased interference from subsequent storage instances on the preceding instance.

In summary, the experiments observed the retroactive inhibition phenomenon in the graph and verified the impact of probe vectors with different $\rho$ values on retrieval effectiveness through awakening based on both upstream and downstream information after storing multiple instances. $\rho$ is negatively correlated with retrieval effectiveness, confirming the influence of the number of iterations in the storage process on retrieval effectiveness.

\subsection{Capacity Testing}\label{sec:capa}
As previously mentioned, different instances stored in the directed graph interfere with each other to some degree, affecting their retrieval. In the capacity testing experiment, a certain number of storage instances were randomly generated, and all were subsequently stored in the same directed graph. Each stored instance was then individually retrieved, and $\alpha$ was calculated. Storage was considered successful if $\alpha\geq90\%$. Retrieval was conducted using the awakening method based on upstream information, which reflects the number of instances that can be successfully and completely stored in a directed graph after storing multiple instances.

\begin{figure}[ht]
    \centering
    \subfigure[]{
        \includegraphics[width=5.5cm]{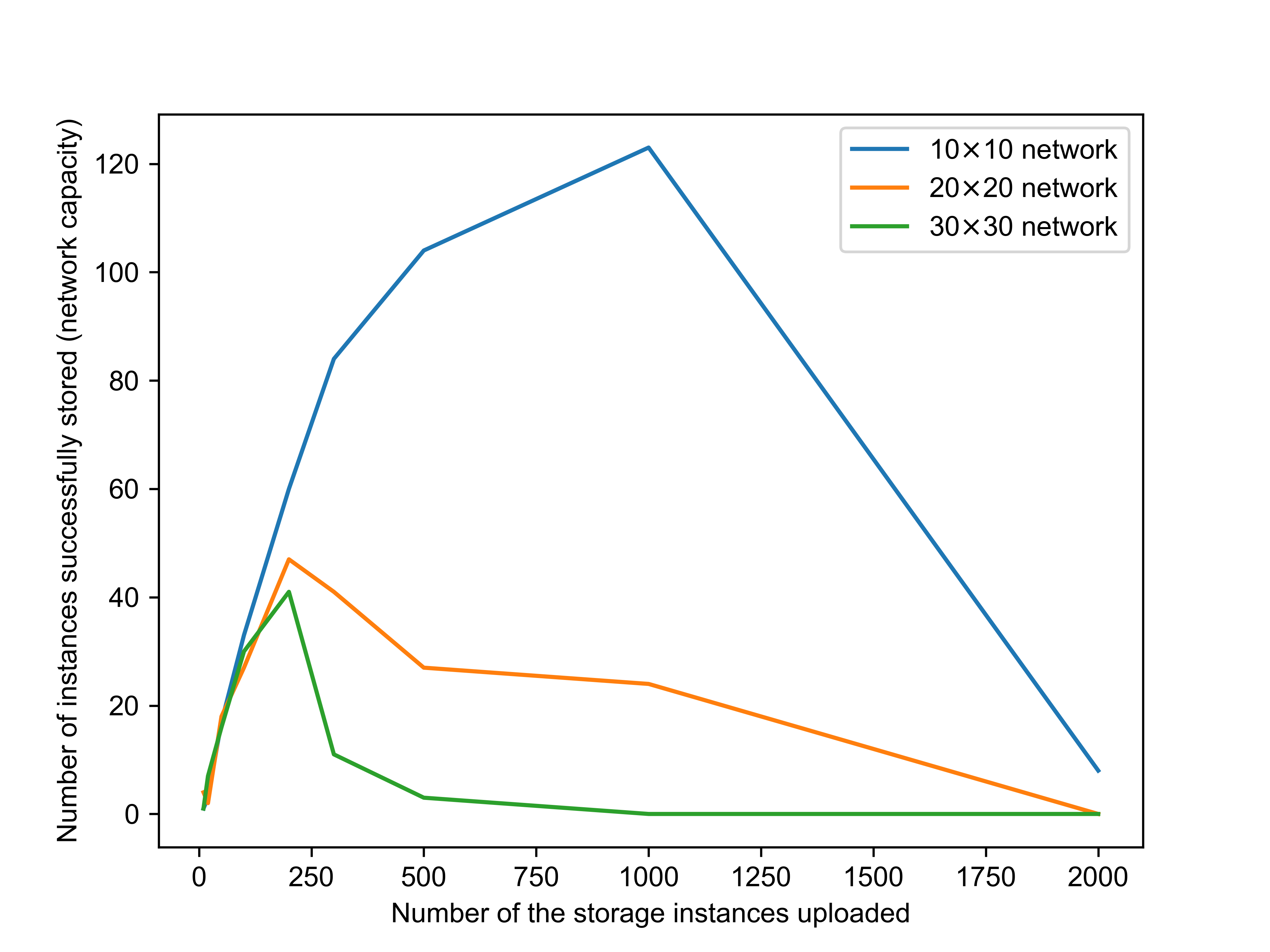}
        \label{fig:basecapa}
    }
    \subfigure[]{
        \includegraphics[width=5.5cm]{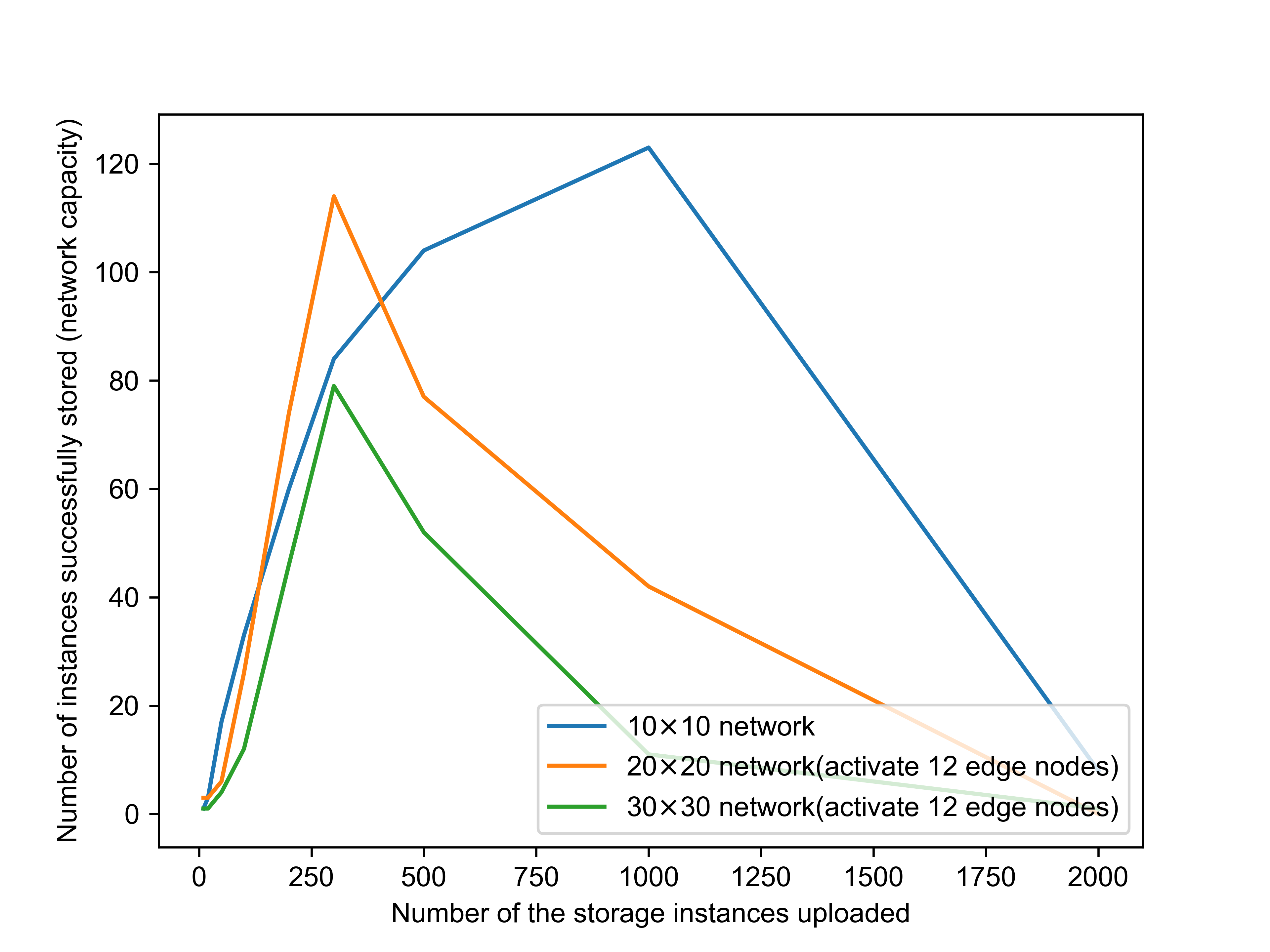}
        \label{fig:activate12capa}
    }
    \subfigure[]{
        \includegraphics[width=5.5cm]{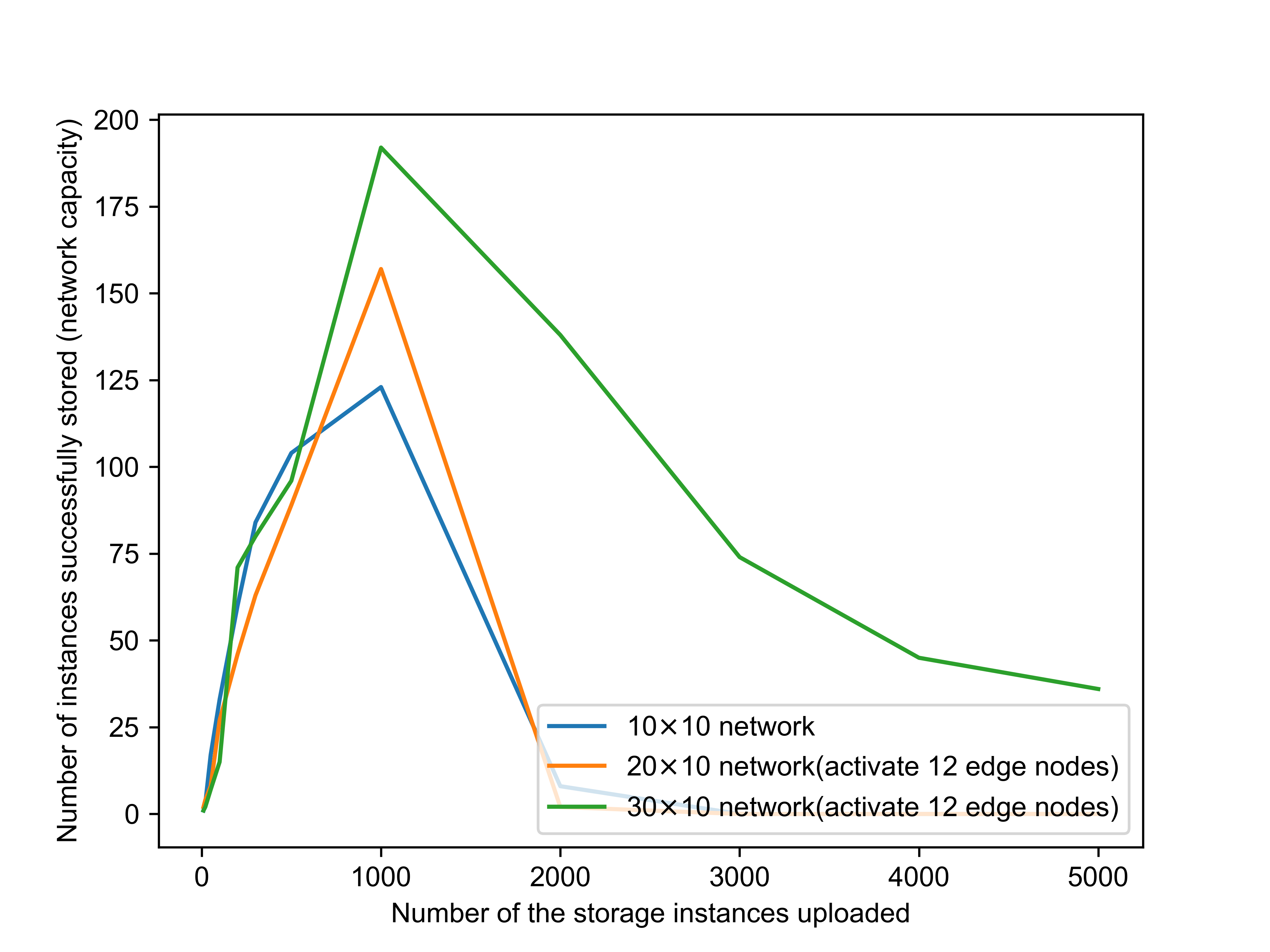}
        \label{fig:depthcapa}
    }
    \caption{(a):Capacity Measured by Individually Retrieving Instances After Storing Different Quantities (b):Capacity Measured After Changing Number of Activated Nodes in Storage Instances (c):Network Capacity Measured with Consistent Network Depth}
\end{figure}

The experiment was conducted in 10 $\times$ 10, 20 $\times$ 20, and 30 $\times$ 30 networks. Several sets of storage instances with 30\% activated edge nodes were randomly generated, with varying numbers in each set. The instances were incrementally stored in the same directed graph, group by group. The learning rate $L_r$ was set at 10\%, and each instance underwent 30 iterations of the storage process. The results, as shown in Figure \ref{fig:basecapa}, indicate that the storage capacity of directed graph networks is limited, and larger networks are not necessarily better; smaller-scale networks seem to have a larger capacity. For the same network, more samples can be stored when internal resource competition is not intense. Once the storage limit is reached, storing more samples can disrupt the paths occupied by previous instances, leading to retrieval failure of instances that were originally remembered.
 
We conducted further testing regarding the issue of larger directed graph networks having smaller capacities. To control variables, the number of activated edge nodes in storage instances randomly generated for 20 $\times$ 20 and 30 $\times$ 30 networks was 12 (10$\times$4$\times$30\%) in both cases, and experiments were conducted in these networks with the same parameters as in Figure \ref{fig:basecapa}. The results, as shown in Figure \ref{fig:activate12capa}, indicate that reducing the number of activated nodes can increase the storage capacity of directed graphs. However, larger networks still have a lesser capacity than smaller-scale networks.
 
In capacity tests based on square directed graphs, the phenomenon of larger network sizes corresponding to smaller capacities was observed. A possible reason for this is that the square network topology, with equal length and width, is not conducive to resource conservation, as the paths formed tend to be deeper. Therefore, we tested networks with a consistent depth but varying widths. Capacity tests were conducted in 20 $\times$ 10 and 30 $\times$ 10 networks, with experimental parameters consistent with those in Figure \ref{fig:activate12capa}. The results, as shown in Figure \ref{fig:depthcapa}, indicate that in networks with consistent depth, larger network sizes correspond to greater capacities. The cerebral cortex has a similar structure, with consistent depth across different areas (six layers of cells), but also has the ability to laterally expand, forming a flat structure.
 
A total of 1000 storage instances with 12 activated edge nodes were randomly generated and uploaded to 30 $\times$ 10, 30 $\times$ 20, and 30 $\times$ 30 directed graphs. The average numbers of nodes and node connections in the paths formed were calculated, to compare the resource consumption of networks with different depths. The experimental results are shown in Table \ref{tab:networkattr}, and indicate that, with the same width, networks with smaller depths consume fewer resources per instance, forming paths with fewer nodes. This explains why the directed graph networks with smaller depths performed better in the earlier experiments. A related biological fact is that the cerebral cortex has relatively few layers, forming a flat structure that is not deep, but that has extensive horizontal expansion capabilities.

\begin{table}[!ht]
    \centering
    \begin{tabular}{O{3cm}|O{5cm}|O{5cm}}
    \toprule
        Size of Directed Graph Network & Number of Connections Between Nodes in Connected Subgraph of Instance & Average Number of Nodes Constituting Path \\ \midrule
        30 × 10 & 364.459 & 54.439 \\ \midrule
        30 × 20 & 813.775 & 83.721 \\ \midrule
        30 × 30 & 1385.789 & 113.686 \\ \bottomrule
    \end{tabular}
    \caption{Attributes of Dominant Paths of Storage Instances Uploaded to Networks of Different Depths}
    \label{tab:networkattr}
\end{table}

\subsection{Capacity Comparison}
We compare the capacity of the proposed model with that of the Hopfield network measured in experiments. Like the criterion of $\alpha\geq90\%$ set in this paper, in the Hopfield network, a retrieved vector is considered successfully remembered if its similarity with the original vector exceeds 90\%. The memory efficiency of the Hopfield network is optimal when each binary bit of a storage instance is independently generated with a 50\% probability of being one of two values  \cite{mceliece_capacity_1987}, which corresponds to randomly activating 50\% of the edge nodes in the proposed model. To control variables, other parameter settings remained the same as in Section 6.5, but we randomly activated 50\% of the edge nodes. At the same time, $\rho$ for the probe vector used in the Hopfield network was the same as $\rho_{max}$. 

\begin{table}[htbp]
\centering
\caption{Capacity of Our Method and Hopfield Network}
\label{table:capacomparison}
\begin{tabular}{c|c|c|c}
\toprule
\multicolumn{2}{c|}{Our Method} & \multicolumn{2}{|c}{Hopfield} \\ \midrule
Size & Measured Capacity & Size & Measured Capacity \\ \midrule
$30 \times 10$ & 108 & $300$ & 20 \\ \midrule
$30 \times 20$ & 32 & $600$ & 22 \\ \midrule
$30 \times 30$ & 15 & $900$ & 25 \\ \bottomrule
\end{tabular}
\end{table}

The experimental results, as shown in Table \ref{table:capacomparison}, are similar to our earlier conclusions. The proposed model shows higher storage efficiency in networks with smaller depths and outperforms the Hopfield network in capacity. However, as the depth increases, the capacity gradually becomes less than that of the Hopfield network. Furthermore, in the experiments, the capacity of the Hopfield network dropped rapidly when the binary bits of the storage instance were unevenly distributed, which corresponds to a decrease in the activation ratio in the proposed method. As mentioned above, reducing the activation ratio in our method can increase the storage capacity. Therefore, our model can flexibly adjust storage capacity by varying the proportion of activated nodes, whereas the Hopfield network cannot. In addition, compared with the fully connected network of the Hopfield model, our model’s sparsely connected network consumes fewer resources and achieves greater storage capacity, especially in networks with smaller depths.

However, our method of retrieval is based on extracting downstream nodes from upstream nodes, and it cannot retrieve information represented by downstream nodes if upstream nodes are lost, meaning it cannot extract upstream nodes based on downstream nodes. Meanwhile, the Hopfield network does not differentiate between upstream and downstream nodes.

\subsection{Capacity Testing of Networks with Six Layers of Depth}
\begin{figure}[htbp]
    \centering
    \includegraphics[width=14cm]{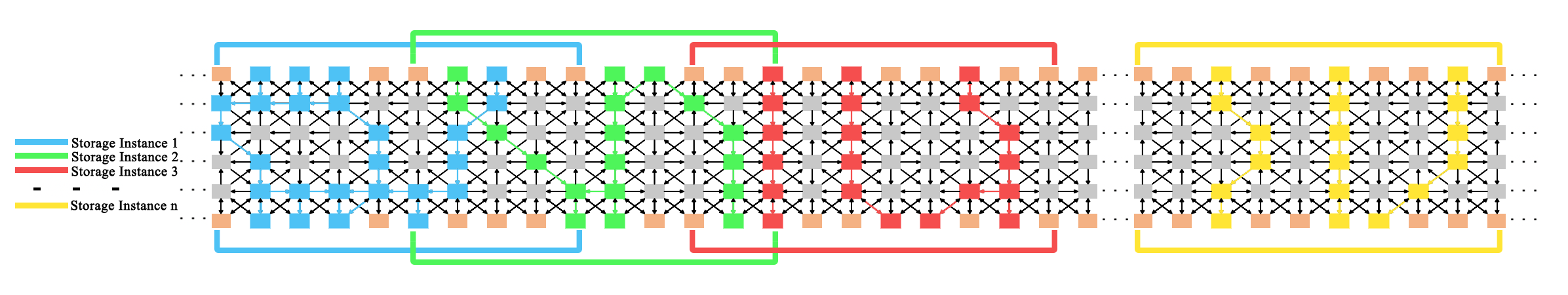}
    \caption{Compactly and Randomly Uploading Memory Instances to Different Positions on Directed Graph}
    \label{fig:rndup}
\end{figure}

The cerebral cortex of higher primates has a vertical depth of six layers of cells and the capability for lateral expansion. Mimicking this form of the brain’s cortex, the depth of the directed graph network was set to six layers. In this configuration, the storage instance retains components A and D, with no B or C. Component A is used for input and D for output, consistent with previous capacity testing experiments, and retrieval is based on upstream information. To control variables, the number of bits in the storage instance vectors was maintained at 20 bits (10 bits each for A and D), with an activation ratio of 50\%. Random starting positions were selected on the top edge of the directed graph, with component A continuously uploaded from the starting position and D uploaded to the symmetric position on the bottom edge, as illustrated in Figure \ref{fig:rndup}. In this case, the storage instances are said to be compactly distributed on the directed graph, with each instance being uploaded on consecutive edge nodes of the graph.
 
\begin{figure}[ht]
    \centering
    \subfigure[Compactly Distribution]{
        \includegraphics[width=6cm]{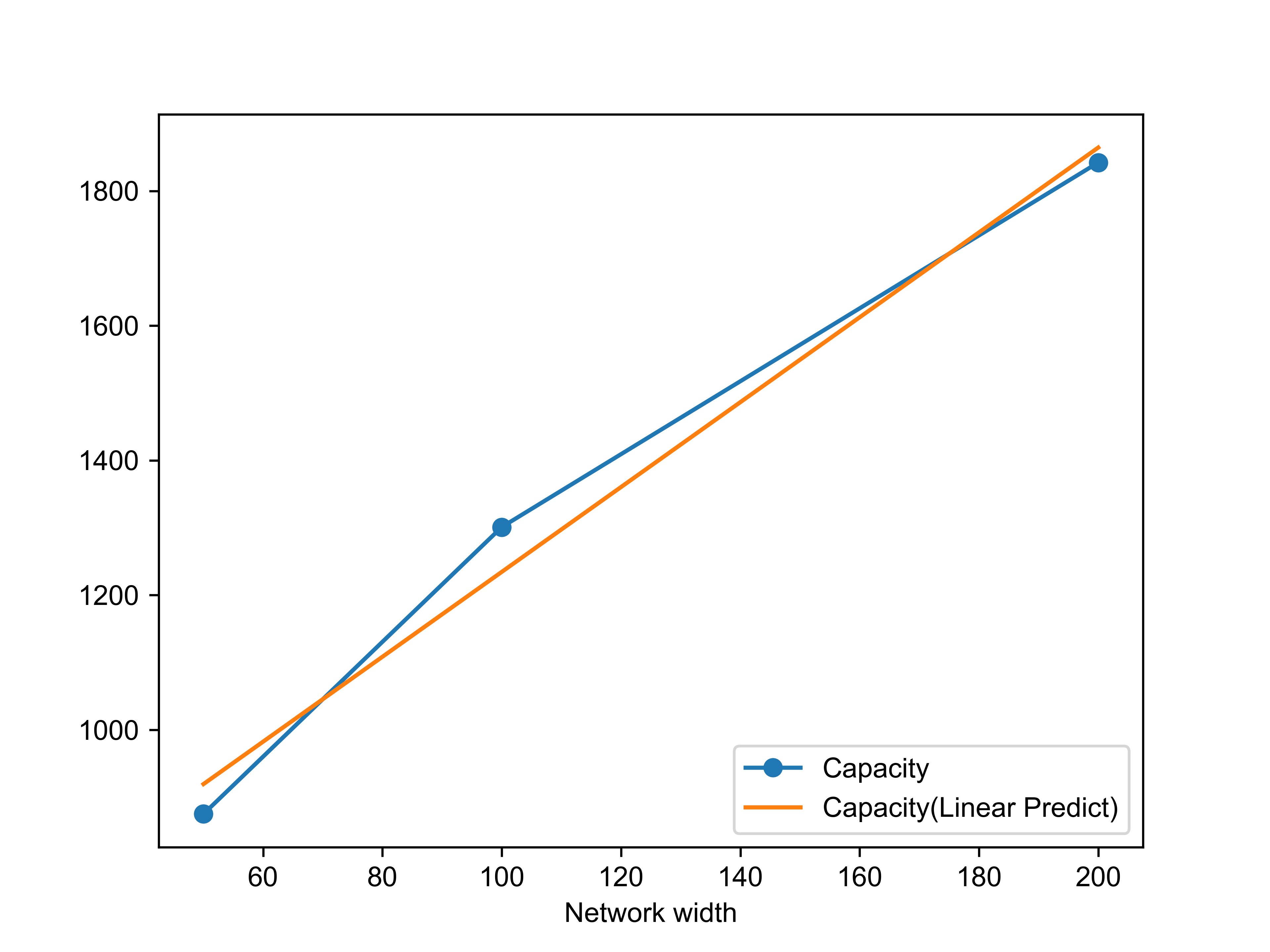}
        \label{fig:comdist}
    }
    \subfigure[Dispersed Distribution]{
        \includegraphics[width=6cm]{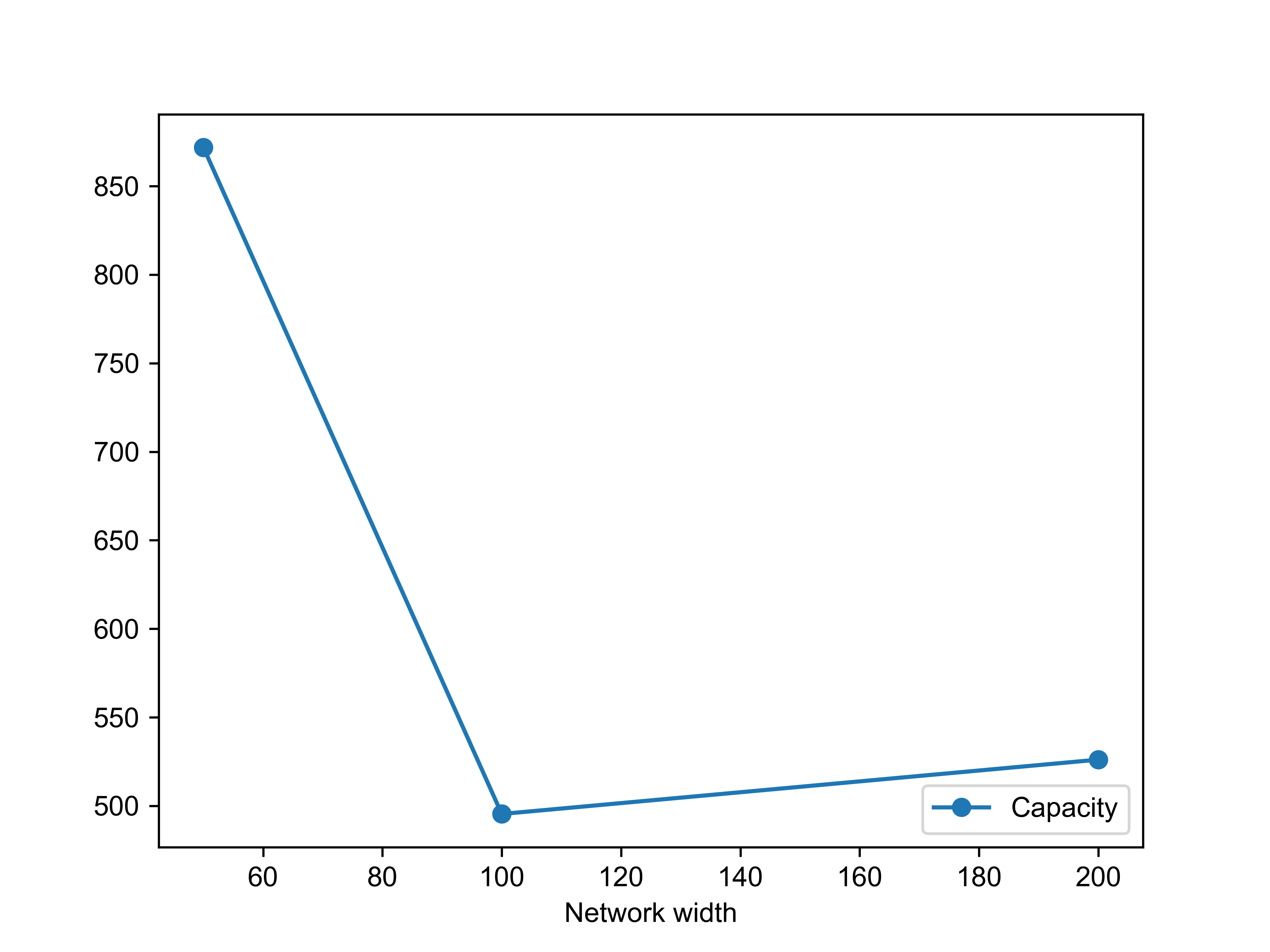}
        \label{fig:disdist}
    }
    \caption{Capacity of Directed Graph Networks with Six Layers of Depth and Different Widths}
\end{figure} 

The experiment was conducted in networks of sizes 50 $\times$ 6, 100 $\times$ 6, and 200 $\times$ 6. Several groups of storage instances were randomly generated, each with varying numbers of instances. Instances were incrementally stored in the same directed graph, group by group. The learning rate $L_r$ was set at 10\%, and each instance underwent 30 iterations of the storage process. The capacity testing rules were the same as in Section \ref{sec:capa}, and the test results were as shown in Figure \ref{fig:comdist}, indicating that as the network width increases, its capacity gradually improves, roughly linearly.

We also experimentally tested the storage capacity of the directed graph when instances were uploaded in a dispersed distribution. In this case, A and D components were no longer continuously uploaded from a certain position to the top and bottom edges. Instead, bits of these components were randomly mapped to different positions on the top and bottom edges. The corresponding nodes of each storage instance were no longer concentrated in a continuous area, to test the storage capacity of the directed graph when instances were stored randomly.

The experimental results of the dispersed distribution, as shown in Figure \ref{fig:disdist}, indicate that except for the network with a width of 50, whose capacity remained largely unchanged compared with the compact distribution, the capacities of the other directed graph networks showed a significant decrease and almost stopped increasing as the network widened.

To explore the reasons for this phenomenon, the resource occupation of individual instances was analyzed. The results, as shown in Table \ref{tab:resource_occupation}, indicate that in the case of a compact distribution, the resource occupation of individual instances does not increase as the network widens. In a dispersed distribution, the horizontal distance between activated nodes is typically greater, requiring more resources to connect, leading to a linear increase in resource occupation per instance with increasing width. The resource occupation in a dispersed distribution at a width of 50 was still relatively close to that of a compact distribution, which may explain why the network capacity in a dispersed distribution remains close to that of a compact distribution at this width. Subsequently, as the gap in resource occupation widens and shows linear growth, the capacity becomes significantly lower than in a compact distribution and almost stops growing with the widening of the network.

\begin{table}[h!]
  \centering
  \caption{Resource Occupation of Single Instances in 6-Layer Depth Directed Graphs with Compact and Dispersed Distribution of Different Widths}
  \label{tab:resource_occupation}
  \begin{tabular}{O{4.5cm}O{4.5cm}O{4.5cm}}
    \toprule
    Network Width & Number of Connections Between Nodes in Connected Subgraph of Instance & Average Number of Nodes Constituting Path \\
    \midrule
    \multicolumn{3}{c}{Compact Distribution} \\
    50  & 50.174 & 35.735 \\
    100 & 50.171 & 35.934 \\
    200 & 50.957 & 35.447 \\
    \midrule
    \multicolumn{3}{c}{Dispersed Distribution} \\
    50  & 63.042 & 53.373 \\
    100 & 90.687 & 73.097 \\
    200 & 131.151 & 101.649 \\
    \bottomrule
  \end{tabular}
\end{table}

\subsection{Chain Awakening Experiment with Long Chains}
\begin{figure}[htbp]
    \centering
    \includegraphics[width=14cm]{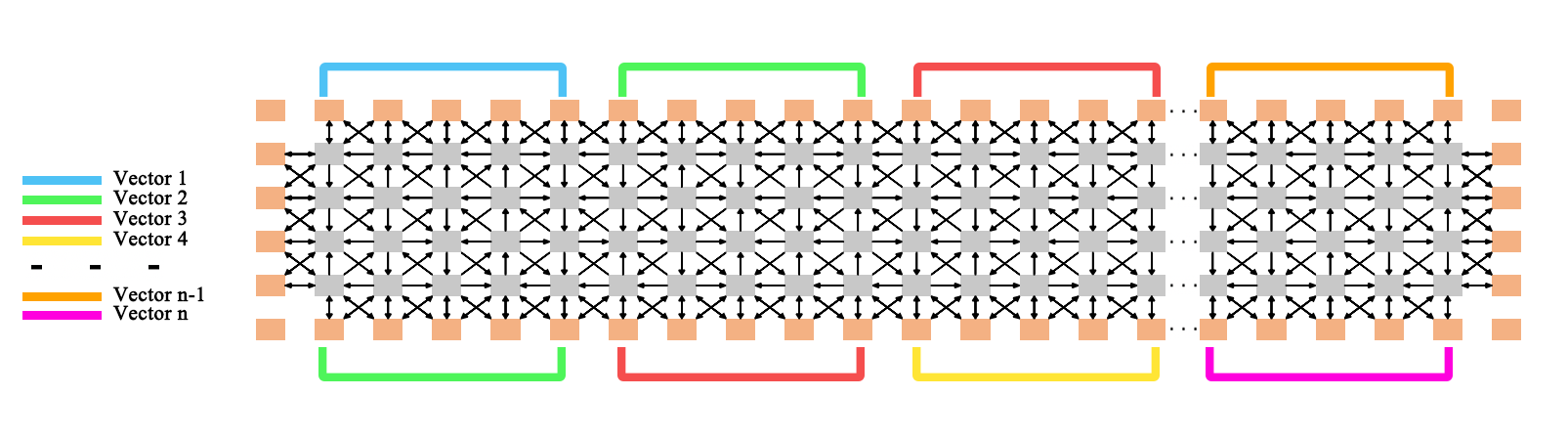}
    \caption{Uploading Pairs of Vectors as Memory Instances to Non-overlapping Positions on Directed Graph}
    \label{fig:chainup}
\end{figure}

Many brain activities, such as sequential working memory, are related to temporal information \cite{xie_geometry_2022}. A long-chain sequential awakening experiment examined a series of instances with sequential associations, where the output component of a memory instance is the input for the following instance, i.e., it acts as an awakening cue. The experiment observed whether such a chain-style awakening can be successful. Multiple vectors of the same length were randomly generated and combined in pairs to form memory instances, with the output vector (D component) of one instance serving as the input vector (A component) of the next. The method of uploading memory instances is similar to Figure \ref{fig:rndup} and illustrated in Figure \ref{fig:chainup}, the edge nodes used by each instance do not overlap, indicating that the instances do not share input or output nodes.
 
\begin{figure}[htbp]
    \centering
    \includegraphics[width=6cm]{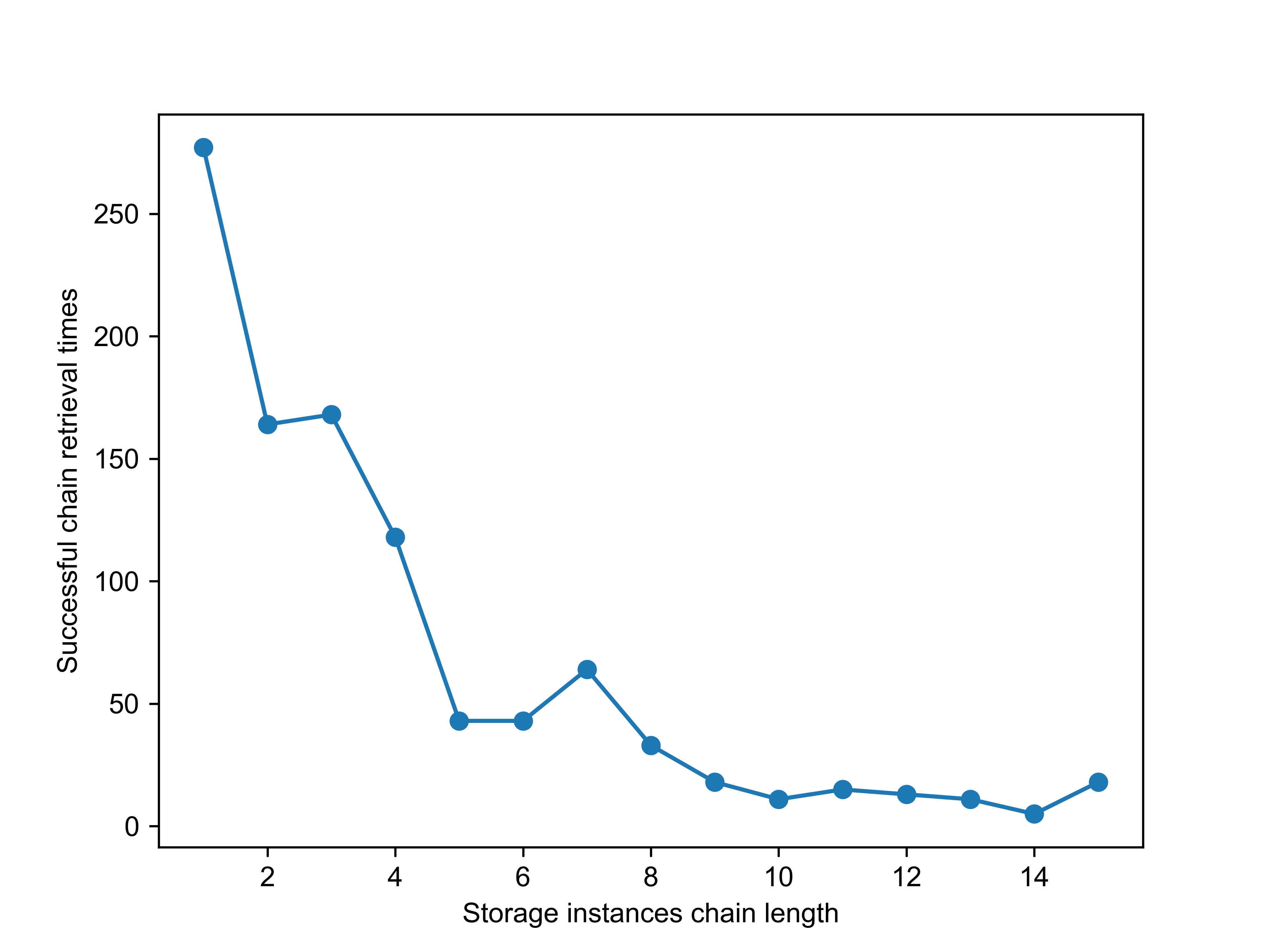}
    \caption{Distribution of Chain Lengths in Chain Awakening Across 1000 Experiments}
    \label{fig:chaindist}
\end{figure}

The experiment was conducted in a 200 $\times$ 6 network. Sixteen vectors of length 10 were randomly generated, with adjacent vectors paired to form storage instances, totaling 15 pairs. In this setup, the output of one pair of instances served as the awakening cue for the next pair. The instances were incrementally stored in the same directed graph in sequential order, with the learning rate $L_r$ set at 10\%, and each instance underwent 30 iterations of the storage process. After storage, each instance was sequentially retrieved in order. Retrieval of the next instance proceeded only if the previous retrieval was successful; otherwise, the retrieval chain was broken, and subsequent instances were not retrieved. In this experiment, $\alpha\geq80\%$ was considered successful storage, and other capacity testing rules were the same as in Section \ref{sec:capa}. The experiment was repeated 1000 times. The test results are shown in Figure \ref{fig:chaindist} and indicate that the model can successfully chain-awaken sequentially stored memory instances, but the probability of successful awakening decreases for instances closer to the end of the chain. This is because the failure to retrieve any instance leads to the failure of all subsequent instances in the chain.

\section{Conclusion}
We simulated the memory mechanism of the cerebral cortex and its storage performance under the constraints of the anatomical and electrophysiological principles of neurobiology. A parallel distributed storage model based on directed graphs and an autonomous path learning algorithm were proposed, offering a new perspective on how biological neural networks achieve memory. Our model provides an explanatory framework to understand the implementation mechanisms of the memory trace theory.

The directed graph model in the algorithm is active, with each node capable of personalized adaptive learning based on local neighborhood information, modeling the memory system of biological neural networks. This breaks away from the reliance of the traditional graph model on a global view for operation, lacking autonomous parallel distributed processing capabilities, and is closer to actual biological neural networks. Inspired by the capability of biological neurons to transmit electrical signals, this adaptive learning behavior was simulated through microcircuits centered on variable resistors, successfully realizing the simulation of the storage and retrieval processes in the entire directed graph network on a computer. In simulations, the model could distribute storage instances across the directed graph, transforming them into connected paths, and achieving path storage and retrieval within the graph. The model’s generalizability was verified, and it was shown to be capable of achieving memory functions in different topological structures of directed graphs. Tests determine the capacity of networks of various scales, verifying factors influencing capacity size, as network scales, edge node activation ratios, and dispersion of storage instances all affect network capacity. Particularly, referencing the structure of the cerebral cortex in higher primates, the storage capabilities of networks with six layers of depth were verified, exploring the relationship between the capacity of six-layer networks and their width.

The process of storing and retrieving instances simulates the biological processes of memory and recall. Experimental results showed certain similarities between the model and biological neural networks. For example, the number of iterations required for a directed graph network to store different memory instances and achieve the same retrieval effectiveness varies, as does the time required for humans to remember different content. Furthermore, our model universally possesses memory functions in directed graphs of different sizes and topological structures, similar to how the microscopic differences in people’s brain networks do not affect the effectiveness of memory functions. Our model performed better and had a larger capacity in networks with smaller depths, with capacity increasing linearly with width, resembling the flat morphology and horizontally expansive cortical structure of the brains of higher primates.

In summary, we proposed a self-learning directed graph model that does not require a global perspective, based on the simulation of biological neural networks. The directed graph achieves storage functionality while decentralized, based on local information and adaptive learning capabilities. The model exhibits biological plausibility, aiding in elucidating the neural activities underlying memory. It also serves as an inspiration for further research into memory neural mechanisms, enriches neural computational models, and offers new perspectives and ideas for neural computation research.

\section*{Declaration of competing interest}
The authors declare that they have no known competing financial interests or personal relationships that could have appeared to influence the work reported in this paper.
\section*{Data availability}
No data was used for the research described in the article.
\section*{Acknowledgments}
The authors gratefully acknowledge the support given by the National Natural Science Foundation of China (Grant No. 61771146).

\bibliographystyle{elsarticle-num}
\bibliography{ref}

\end{document}